\let\accentvec\vec
\documentclass{aa}
\usepackage[latin1]{inputenc}
\usepackage[T1]{fontenc}
\usepackage{lmodern} 
\usepackage{graphicx}
\usepackage{subfig}
\usepackage[font=small,labelfont=bf,tableposition=top]{caption}
\DeclareCaptionLabelFormat{andtable}{#1~#2  \&  \tablename~\thetable}
\usepackage{float}
\usepackage{tabu}
\usepackage[squaren, Gray, cdot]{SIunits}
\usepackage[varg]{txfonts}
\usepackage{textcomp} 
\usepackage{color}

\let\vec\accentvec

\usepackage {amssymb, mathrsfs} 
\usepackage{multirow}

\usepackage{natbib}
\usepackage[]{longtable} 
\bibpunct{(}{)}{;}{a}{}{,}

\makeatletter

\begin{document} 
\defcitealias{biernauxetal2016}{B16}

\title{Analysis of luminosity distributions of strong lensing galaxies: subtraction of diffuse lensed signal}
\author{J. Biernaux\inst{\ref{inst1}} \and P. Magain\inst{\ref{inst1}} \and C. Hauret\inst{\ref{inst1}}}
\institute{Universit\'e de Li\`ege \\ STAR Research Unit - OrCA Group \\ Quartier Agora. B\^at. B5c \\ All\'ee du 6 Ao\^ut, 19c\\ B-4000 Li\`ege 1 (Sart-Tilman)\\ Belgique\\ jbiernaux@ulg.ac.be \label{inst1} \\ pierre.magain@ulg.ac.be \label{inst2}\\ clementine.hauret@ulg.ac.be \label{inst3}} 
\date{May 2017}
\abstract {Strong gravitational lensing gives access to the total mass distribution of galaxies. It can unveil a great deal of information about the lenses' dark matter content when combined with the study of the lenses' light profile. However, gravitational lensing galaxies, by definition, appear surrounded by lensed signal, both point-like and diffuse, that is irrelevant to the lens flux. Therefore, the observer is most often restricted to studying the innermost portions of the galaxy, where classical fitting methods show some instabilities.}
{We aim at subtracting that lensed signal and at characterising some lenses' light profile by computing their shape parameters (half-light radius, ellipticity, and position angle). Our objective is to evaluate the total integrated flux in an aperture the size of the Einstein ring in order to obtain a robust estimate of the quantity of ordinary (luminous) matter in each system.}   
{We are expanding the work we started in a previous paper that consisted in subtracting point-like lensed images and in independently measuring each shape parameter. We improve it by designing a subtraction of the diffuse lensed signal, based only on one simple hypothesis of symmetry. We apply it to the cases where it proves to be necessary. This extra step improves our study of the shape parameters and we refine it even more by upgrading our half-light radius measurement method. We also calculate the impact of our specific image processing on the error bars.}
{The diffuse lensed signal subtraction makes it possible to study a larger portion of relevant galactic flux, as the radius of the fitting region increases by on average 17\%. We retrieve new half-light radii values that are on average 11\% smaller than in our previous work, although the uncertainties overlap in most cases. This shows that not taking the diffuse lensed signal into account may lead to a significant overestimate of the half-light radius. We are also able to measure the flux within the Einstein radius and to compute secure error bars to all of our results.}
{}
\titlerunning{Subtraction of lensed signal}

\maketitle

%%%%%%%%%%%%%%%%%%%%%%%%%%%%%%%%%%%%%%%%%%%%%%
\section{Introduction}\label{sec_intro}

\indent Gravitational lensing is a powerful tool to address many astrophysical questions, ranging from cosmology to galaxy formation and evolution, as it offers a precise technique for studying mass distributions in general. Strong lensing, in particular, gives access to valuable information about galactic mass profiles. In the case of early-type galaxies, as the study of their dynamics is rather challenging, especially at high-$z$ \citep{Rom2003, Cappellari2015}, strong lensing practically constitutes our only proxy for the mass distributions. Luckily, thanks to their usually higher surface mass density, elliptical galaxies are more often involved in gravitational lensing phenomena than spirals. While strong lensing offers a possibility of understanding their total mass distribution, comparing it to their brightness profile could provide valuable information about dark matter in early-type galaxies \citep{Bertinetal1994, Rom2003, Dekeletal2005, Cappellari2015}. It is therefore of highest interest to accurately determine the luminosity distribution of lensing galaxies. \\

\indent However, the existence of lensed images around a galaxy makes the study of its light profile troublesome. Indeed, the deflected light from the background source produces point-like and diffuse components that act as a parasite signal when one wants to access only the lens brightness. The authors address this issue in \citet{biernauxetal2016} (hereafter \citetalias{biernauxetal2016}) by implementing a careful subtraction of the point-like lensed images for seven gravitational lensing systems from the CfA-Arizona Space Telescope LEns Survey (CASTLES) database \citep{Castles}. A precise point-spread function (PSF) determination is necessary for achieving a good point sources subtraction. Like many authors who have processed lenses in the framework of the COSmological MOnitoring of GRAvItational Lenses (COSMOGRAIL) project \citep{Chantryetal2010, Courbinetal2011, Sluseetal2012a}, we use the MCS deconvolution algorithm \citep{Magainetal1998, ChantryMag2007}. It is well suited to gravitational lensing images, as it consists in iteratively subtracting a diffuse component, including any non-point-like object, such as galaxies and lensed arcs, until convergence to an image of the point sources. Moreover, it has the important advantage of not violating the sampling theorem. In \citetalias{biernauxetal2016} as well as in \citet{ChantryMag2007} or \citet{Chantryetal2010}, we explain that other PSF determinations, using, for example, the TinyTim software \citep{TinyTim1, TinyTim2}, are not accurate enough to perform such a subtraction.\\

\indent Nonetheless, even after PSF subtraction, the uncertainties in the outer regions of the galaxy restrain the modelling of the light profile to its inner regions. It is shown in \citetalias{biernauxetal2016} that classical fitting techniques such as GALFIT \citep{galfit1, galfit2} perform poorly in these conditions. We also tackle that problem in \citetalias{biernauxetal2016} by implementing an innovative study of the shape parameters of lensing galaxies, that is, the position angle (PA) of the galaxy major axis, its ellipticity ($\varepsilon$), and its half-light radius ($r_{\rm{eff}}$). We design a method that is not too sensitive to the above-mentioned artefacts: we propose to study each of them individually, so as to avoid the existence of local minima in the parameters space. The basis of this modelling is the computation of isophotes. We show in \citetalias{biernauxetal2016}, from simulations, that our $r_{\rm{eff}}$ measurement technique, which we call the linear regression method (LRM), is more trustworthy than GALFIT in the case of lensing galaxies. The latter shows instabilities regarding the fitting region, but also crucial aspects of image processing, such as the PSF or the signal-to-noise ratio (S/N). We also show that our technique is better-suited for galaxies that are not much larger than the PSF, which comes in handy for lensing images, as we often deal with only the inner parts of the lens. Moreover, the LRM depends less on the value of the S\'ersic index of the model we choose to represent the galaxy. We thus obtain reliable $r_{\rm{eff}}$ values and present all the results in a CDS (Centre de Donn\'ees astronomiques de Strasbourg) table\footnote{http://vizier.cfa.harvard.edu/viz-bin/VizieR?-source=J/A+A/585/A84} \citep{B16CDS} associated to \citetalias{biernauxetal2016}. \\

\indent In this work, we expand the study of the \citetalias{biernauxetal2016} sample. Despite the point-source subtraction, some diffuse lensed signal, in the form of arcs or a ring surrounding the lenses, remains as a nuisance to the study of their light profile. We aim at designing a method to subtract that signal, in order to be able to encompass more of the outer parts of the galaxy in the fitting region. As the value of $r_{\rm{eff}}$ is sensitive to the wings of the galaxy light profile, this constitutes an improvement to the \citetalias{biernauxetal2016} $r_{\rm{eff}}$. We also improve the LRM by switching to a $\chi^{2}$ validity criterion for the linear regression. Since the initial motivation of this work is to compare light distributions to total mass distributions of lensing galaxies, we finally compute the integrated flux within the Einstein radii of our lenses.  \\

\indent This paper is structured as follows. In Sect. \ref{sec_data}, we give some information about our sample. In Sect. \ref{sec_meth}, we explain the additions and improvements to the \citetalias{biernauxetal2016} shape parameters measurement methods that we implement. We discuss the errors in Sect. \ref{sec_err} and eventually apply it to our sample, discussing the results in Sect. \ref{sec_results}. We give a short conclusion in Sect. \ref{sec_ccl}.
%%%%%%%%%%%%%%%%%%%%%%%%%%%%%%%%%%%%%%%%%%%%%%
\section{Data}\label{sec_data}
\indent We are processing the same data as in \citetalias{biernauxetal2016}, where a complete and illustrated description of our sample can be found. In short, our dataset is a seven-systems subsample from the data processed in \cite{Chantryetal2010} and \cite{Sluseetal2012a}, from the CASTLES database \citep{Castles}. Their selection criteria were that the systems should have securely known redshifts for both the lens and the background quasar and they excluded systems with multiple lenses of similar luminosity. As in \citetalias{biernauxetal2016}, we are focusing on quadruply-imaged sources, thus reducing the full sample to our seven systems. The images were obtained with the NIC2 camera of the Near Infrared Camera and Multi-Object Spectrometer (NICMOS) instrument onboard the Hubble Space Telescope (HST) between 1997 and 2004 in the near infrared H-band. The angular scale of these images is 0.075 arcseconds per pixel. Table \ref{table_data} summarises the systems coordinates and redshifts. We use astrometrical results from \cite{Chantryetal2010} and \cite{Sluseetal2012a}, as well as their accurate PSFs.
\begin{table*}[!pht]
\caption{List of the systems that have been processed in \citetalias{biernauxetal2016} and in this work.}
\centering
\begin{tabu}{c  c  c  c  c  c}
\hline
\textit{System} & \textit{\# frames} & \textit{Source redshift} & \textit{Lens redshift} & \textit{RA (J2000)} & \textit{DEC (J2000)} \\ 
\hline
 & & & & & \\
  	  MG0414+0534 & 13 & $ 2.64$ & $0.96$ & 04:14:37.73 & +05:34:44.3\\
	  HE0435-1223 & 4 & $ 1.689$ & $0.46$ & 04:38:14.9 & $-$12:17:14.4\\
	  RXJ0911+0551 & 4 & $2.80$ & $0.77$ & 09:11:27.50 & +05:50:52.0\\
	  SDSS0924+0219 & 8 & $ 1.524$ & $0.359$ & 09:24:55.87 & +02:19:24.9 \\
	  PG1115+080 & 4 & $1.72$ & $0.351$ & 11:18:17.00 & +07:45:57.7\\
	  SDSS1138+0314 & 4 & $ 2.44$ & $0.45 $ & 11:38:03.70 & +03:14:58.0  \\
	  B1422+231 & 4 & $3.62$ & $0.354$ &14:24:38.09 & +22:56:00.6 \\
& & & & & \\
\hline
\end{tabu}
\label{table_data}
\end{table*}   
%%%%%%%%%%%%%%%%%%%%%%%%%%%%%%%%%%%%%%%%%%%%%%
\section{Methods}\label{sec_meth}

\subsection{Pre-processing}\label{sec_pp}

\indent A complete walk-through of the pre-processing can be found in \citetalias{biernauxetal2016}. To sum up, after the common poor quality pixels cleanup and the sky subtraction, a point source subtraction has been conducted on the data frames. Indeed, with the purpose of disentangling signal from the lensed point-like components and from the galaxy, that subtraction consists in building an image of the four point sources, taking the PSF into account, and subtracting it directly from each data frame.

\subsection{Subtracting the diffuse lensed component}\label{sec_arcsub}

\indent The first aim of this work is to extend the lensed signal subtraction to the arc. Similarly to the subtraction of the point-like images, we aim at building an image of the arc and at directly subtracting it from the original data frame. We formulate only one simple hypothesis: because the arc consists of an image of the background galaxy, its light distribution displays, to the first order, the same properties of radial symmetry. In other words, along a galactocentric radius, the arc should have a symmetric light distribution on either side of its maximum intensity.\\ 
\begin{figure*}[!pht]
\caption{Left: sketch of the division of an image of HE0435-1223 into sectors along the angular coordinates. Right: collapsed radial profile of a sector from HE0435-1223. The crosses represent the measured intensity along the radial coordinate. The solid line shows the best-fitting de Vaucouleurs profile regarding the residuals maximum symmetry criterion.}
\centering
\begin{tabular}{c c}
\includegraphics[scale=0.4]{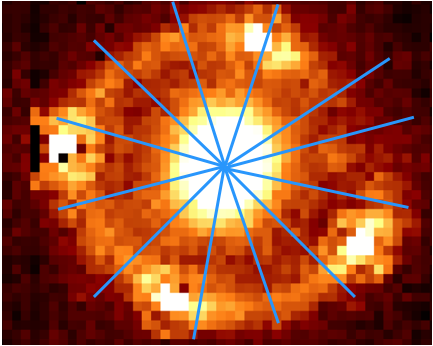} & \includegraphics[scale=0.45]{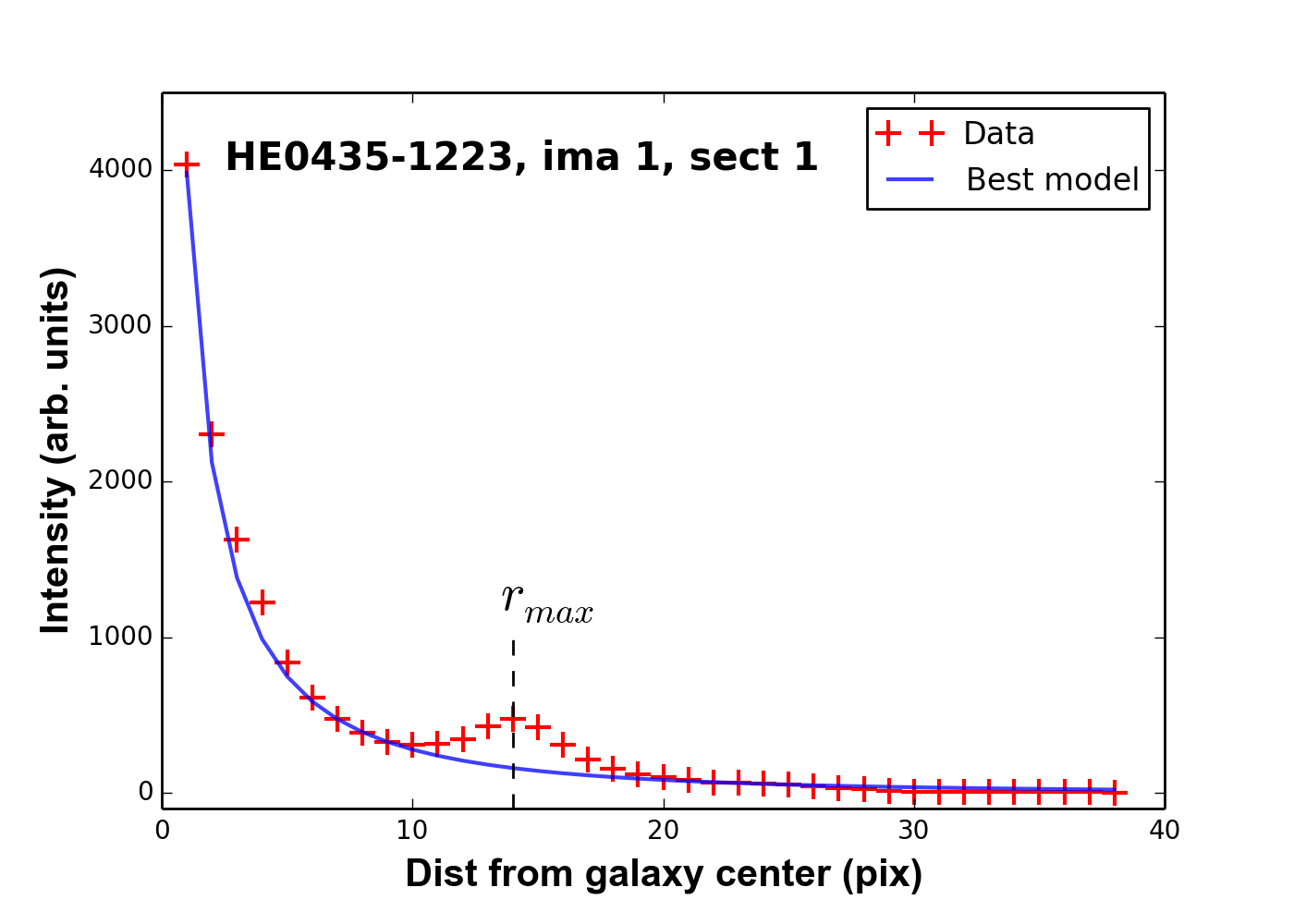}
\end{tabular}
\label{fig_sect}
\end{figure*}

\indent In order to compute the arc radial light distribution, we first locate the maximum intensity of the arc with respect to the centre of our lens. We choose to work in a galactocentric coordinate system. For a few azimutal coordinates $\theta$ (typically about 200, evenly distributed around the whole ring), we locate the pixel with the highest intensity in the arc. This yields an ensemble of $(r_{\rm{max}}~;~\theta)$ datapoints, $r_{\rm{max}}$ being the distance between that pixel and the galactic centre. To refine the location of $r_{\rm{max}}$, we fit a constant plus a linear combination of sines up to the second order on these datapoints:
\begin{equation}
r_{\rm{max}}(\theta) = r^{0}_{\rm{max}}+ a_{\rm{1}}sin(\theta)+a_{\rm{2}}sin(\theta^{2}),
\label{eq_rhotheta}
\end{equation} 
where $r^{0}_{\rm{max}}$ is the average of $r_{\rm{max}}$. We actually conduct this operation separately on pieces of the arc between pairs of point sources. We exclude coordinates where PSF subtraction has been performed. We thus get a function $r_{\rm{max}}(\theta)$ that locates the centre of symmetry of the arc light profile at any angle. We divide the arc image into sectors of various sizes along the angular coordinate, as shown in the left panel of Fig. \ref{fig_sect}. Within each sector, we require $r_{\rm{max}}$ to not vary more than a few percent. However, should $r_{\rm{max}}(\theta)$ be nearly constant, the maximum angular width of a sector is set to a value chosen by the user, typically around 20 degrees. The width of sectors is therefore a parameter of the arc subtraction process, but its influence on the arc subtraction is negligible compared to the other sources of systematics considered in our error calculation (Sect. \ref{sec_err}).\\

\indent The next step consists in collapsing the 2-D image of the arc into 1-D radial profiles. This is done by re-scaling the length of all the rows of pixels of the same angular coordinate, also called traces, within each sector so that they get the exact same $r_{\rm{max}}$. Then, the re-scaled traces are summed and averaged. The resulting radial profile, shown in the right panel of Fig. \ref{fig_sect}, corresponds to the sum of the galaxy and the arc radial profiles, the latter showing up as a "bump" around $r_{\rm{max}}$. We model the galaxy profile $I(r)$ as a de Vaucouleurs law,
\begin{center}
\begin{equation}
I(r) = I_{\rm{0}}\exp{\left(-k \left(\frac{r}{r_{\rm{eff}}}\right)^{1/4}\right)},
\label{eq_dVc}
\end{equation}
\end{center}
where $I_{\rm{0}}$ is the central pixel intensity, $r_{\rm{eff}}$ the half-light radius and $k$ a normalisation constant. This expression is valid for a centred light distribution, which is the case in our galactocentric coordinate system. It is also valid for a circular galaxy. However, we consider the ellipticity of the lens by using $r = \sqrt{(ab})$ as a radial coordinate, where $a$ and $b$ are respectively the semi-major and semi-minor axes. We use as a validity criterion that the de Vaucouleurs profile parameters maximise the symmetry of the residuals in a given region of interest around $r_{\rm{max}}$. The solid curve on Fig. \ref{fig_sect} gives an example of the galactic profile giving the most symmetric residuals wings around $r_{\rm{max}}$. Those residuals act as datapoints for the arc radial light profile. The size of the region of interest is also a parameter of the process: the larger the better, but not so large that we mistakenly consider parts of the image where there should be no arc.\\

\begin{figure*}[!pht]
\caption{Two examples of numerical radial profile computation for one sector of the arc of HE0435-1223 and from SDSS1138+0314. The stars picture the datapoints for the arc radial light profile. The crosses represent the best fitting de Vaucouleurs models for each wing. The solid line shows the linear combination that acts as the arc numerical profile.}
\centering
\begin{tabular}{c c}
\includegraphics[scale=0.45]{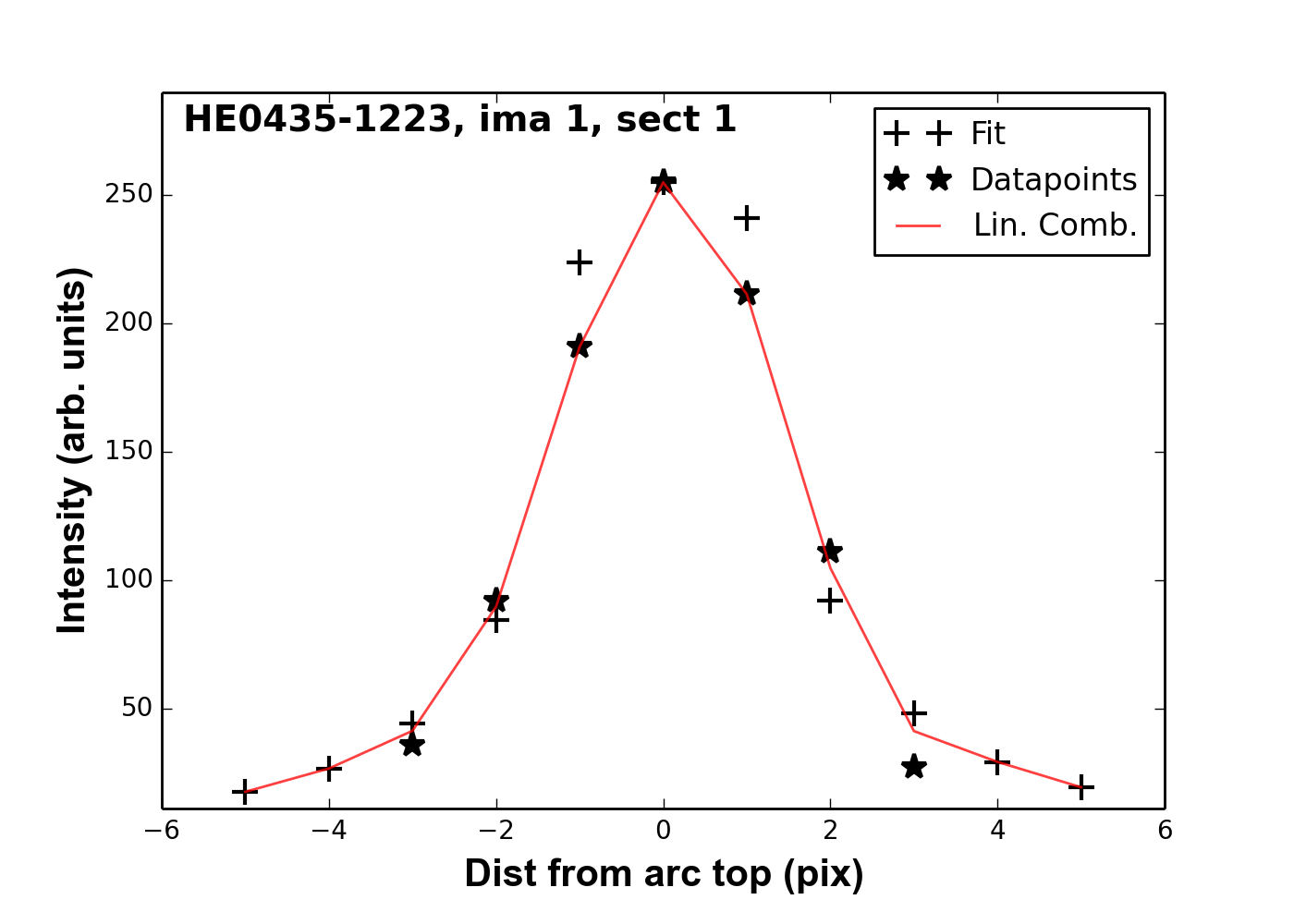} & \includegraphics[scale=0.45]{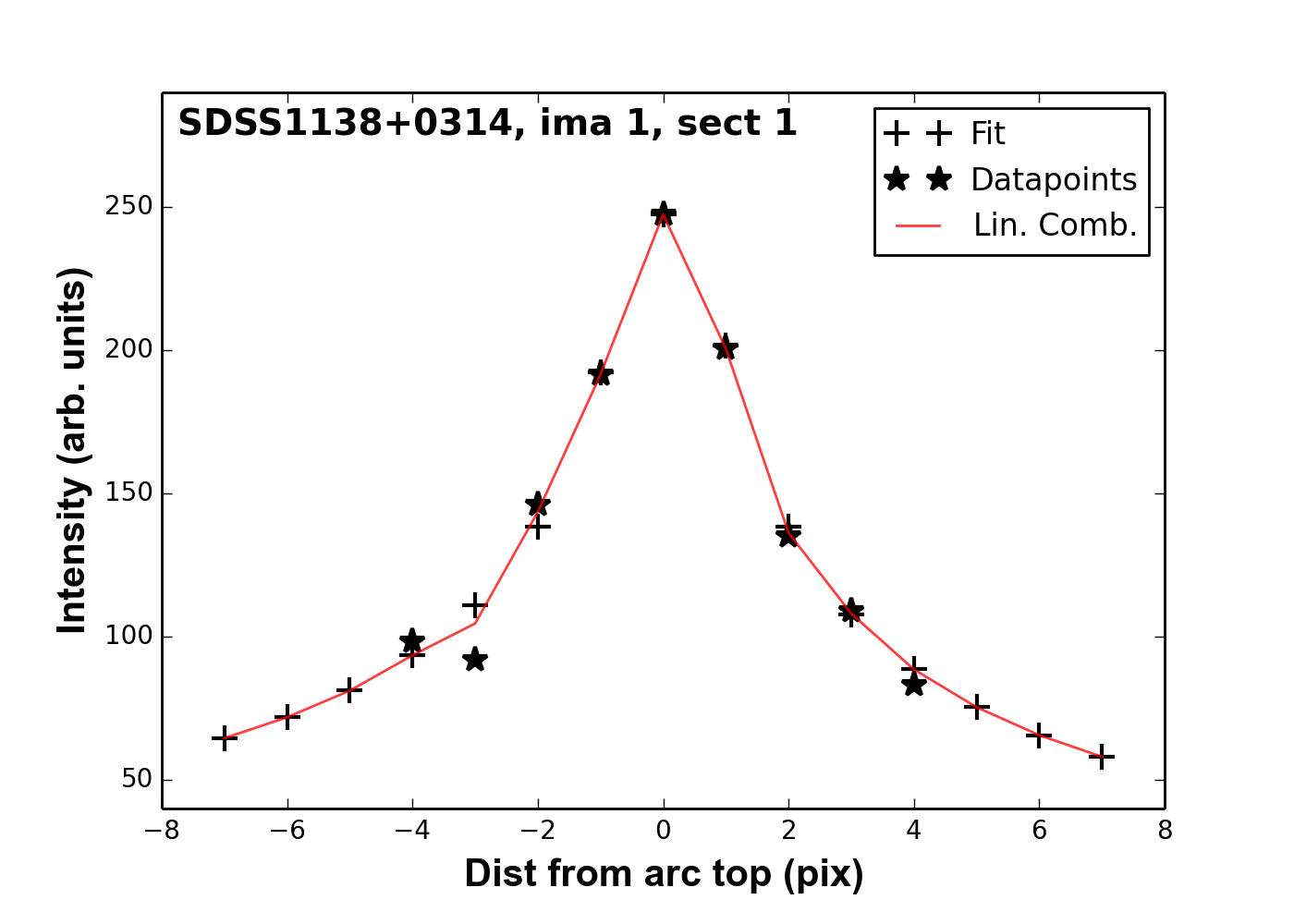}\\[0.5cm]
\end{tabular}
\label{fig_wings}
\end{figure*}

\begin{figure*}[!pht]
\caption{Left: HST-NICMOS data frames from our samples. Right: resulting images after the lensed signal subtraction, point sources and arcs. Only one data frame is shown for each system and only for the systems on which the arc subtraction has been conducted.} % \textit{From top to bottom}: HE0435-1223, SDSS0924+0219, PG1115+080, SDSS1138+0314.
\centering
\begin{tabular}{c c}
& \\
\includegraphics[scale=0.2125]{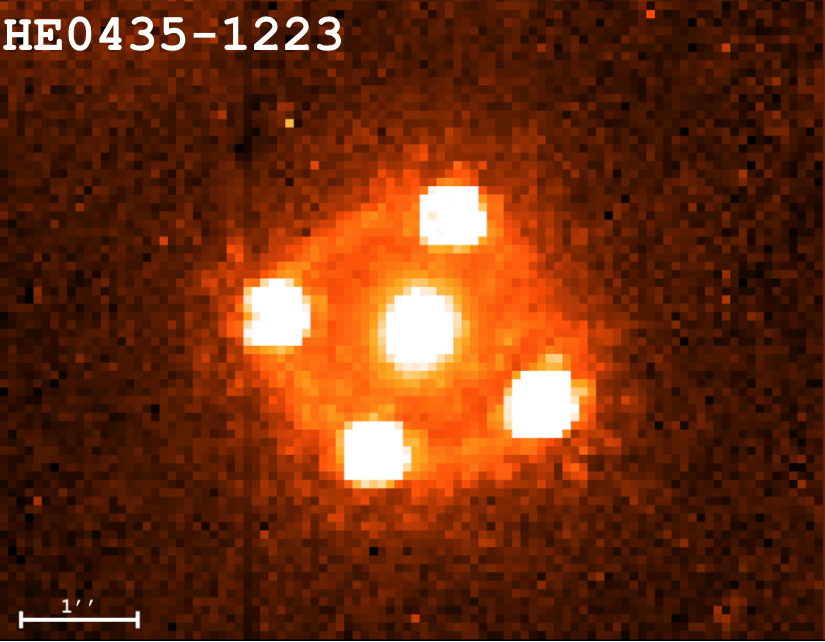} & \includegraphics[scale=0.15]{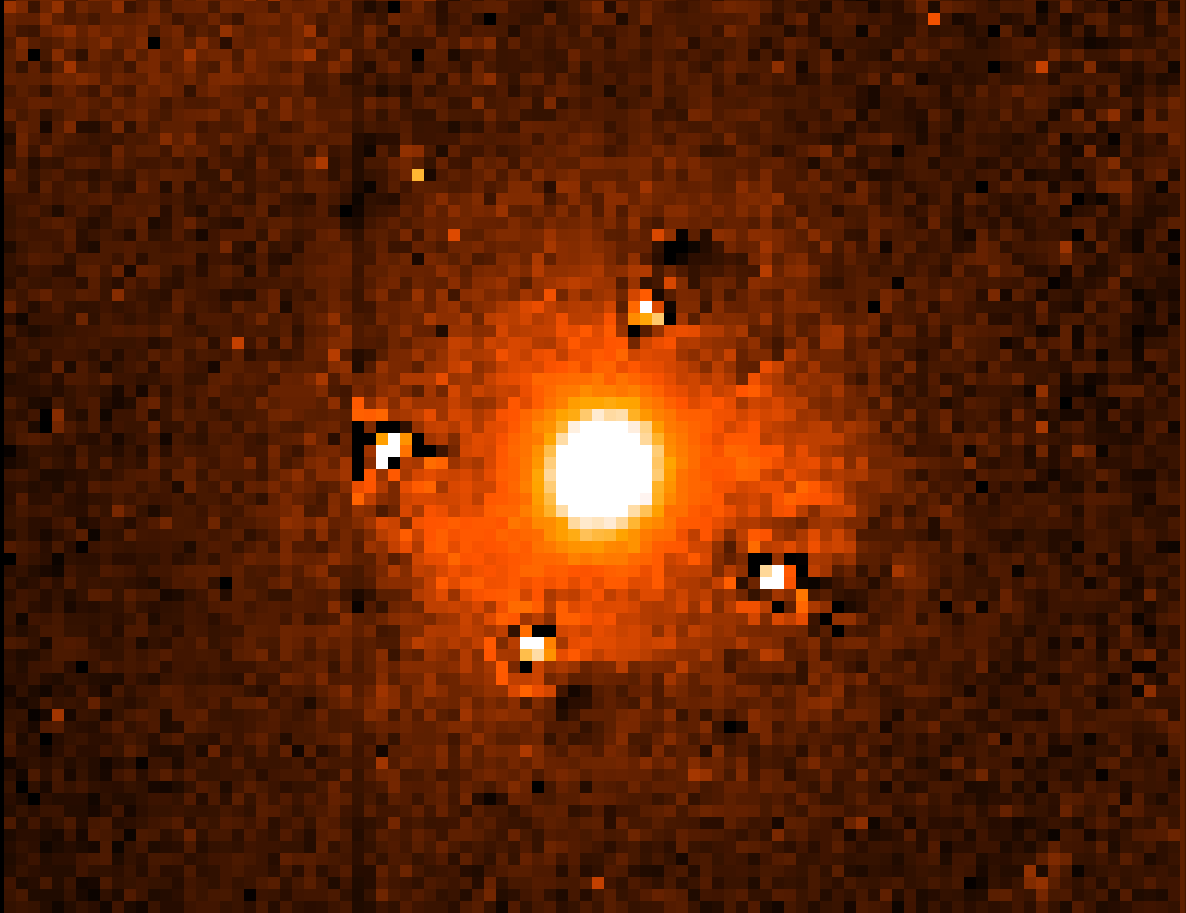} \\[0.25cm]
\includegraphics[scale=0.21]{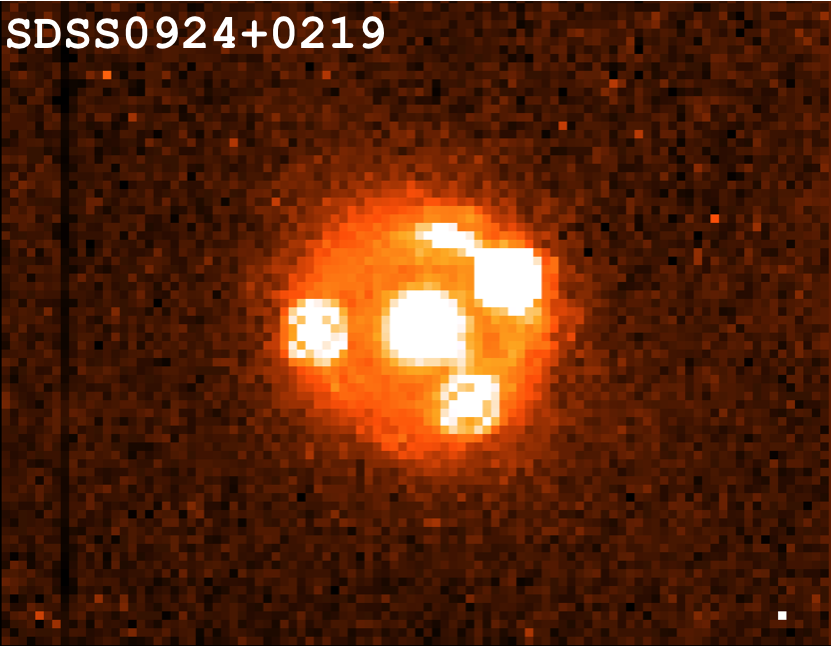} & \includegraphics[scale=0.15]{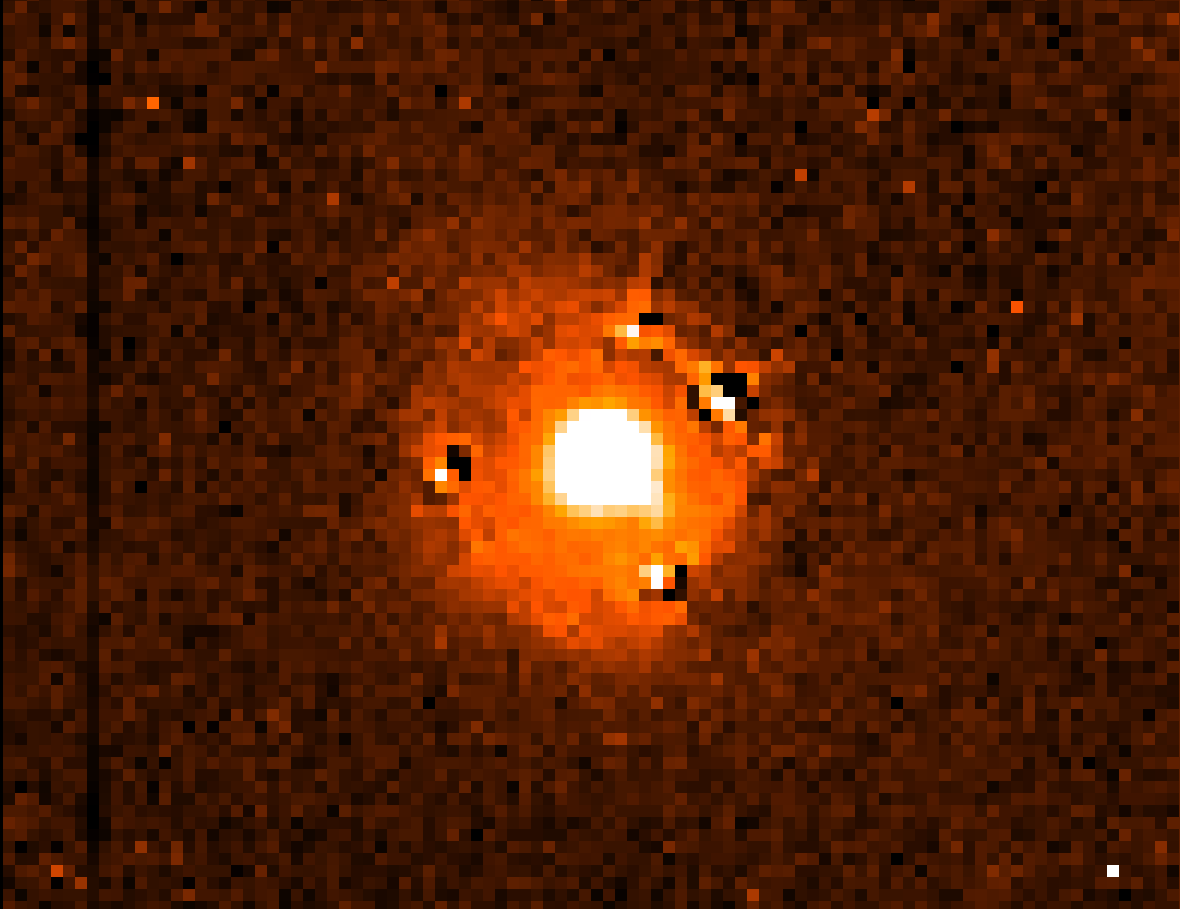} \\[0.25cm]
\includegraphics[scale=0.2175]{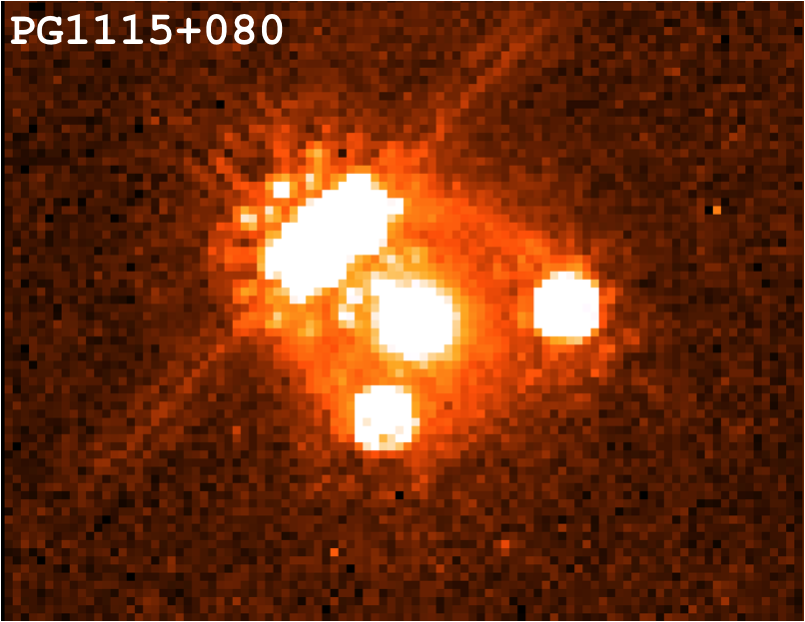} & \includegraphics[scale=0.15]{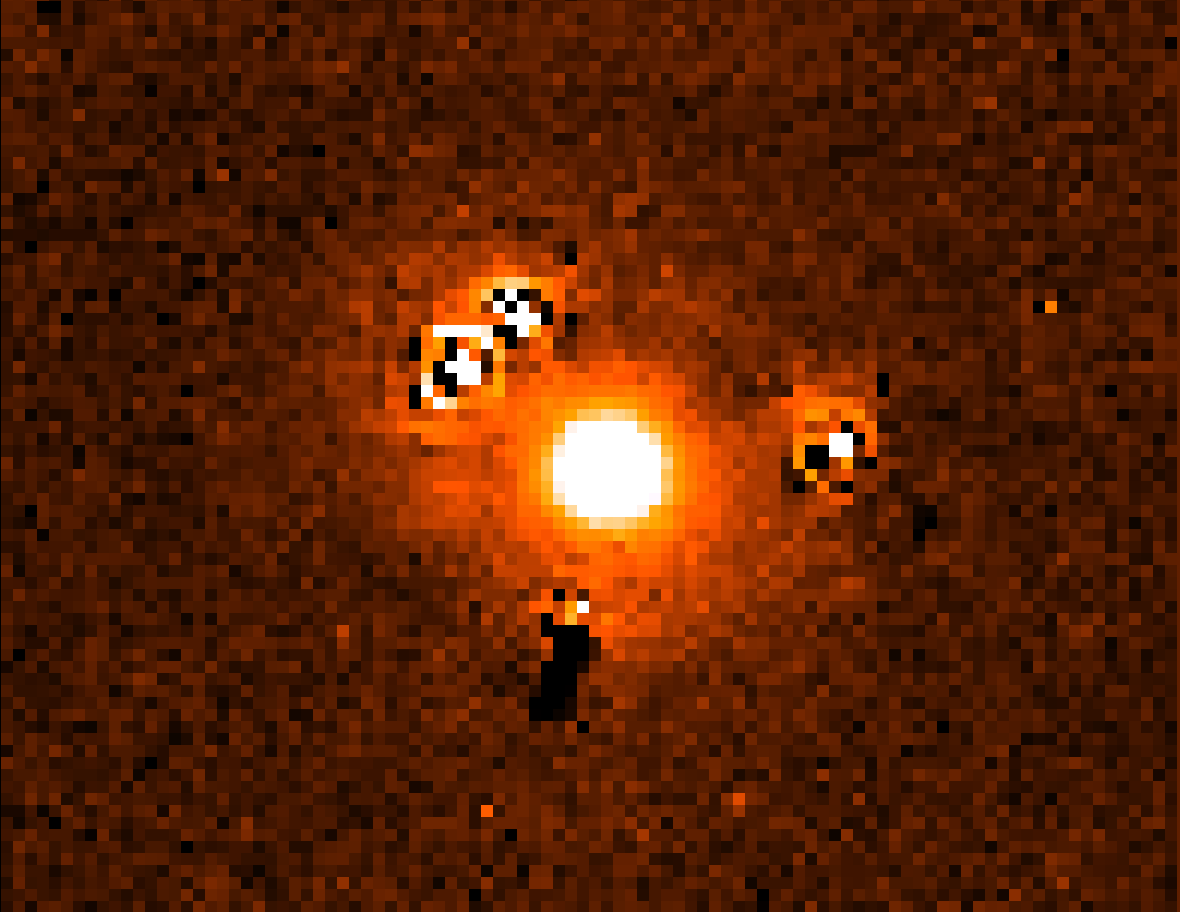} \\[0.25cm]
\includegraphics[scale=0.22]{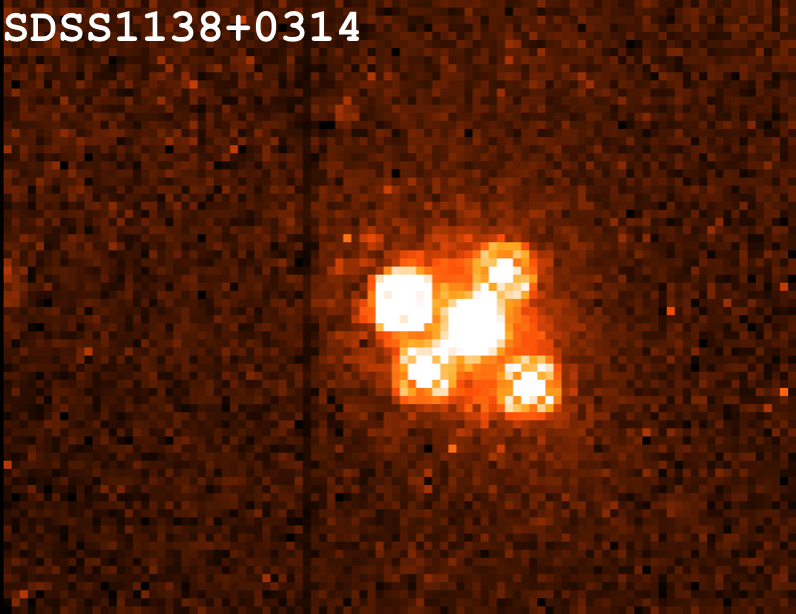} & \includegraphics[scale=0.15]{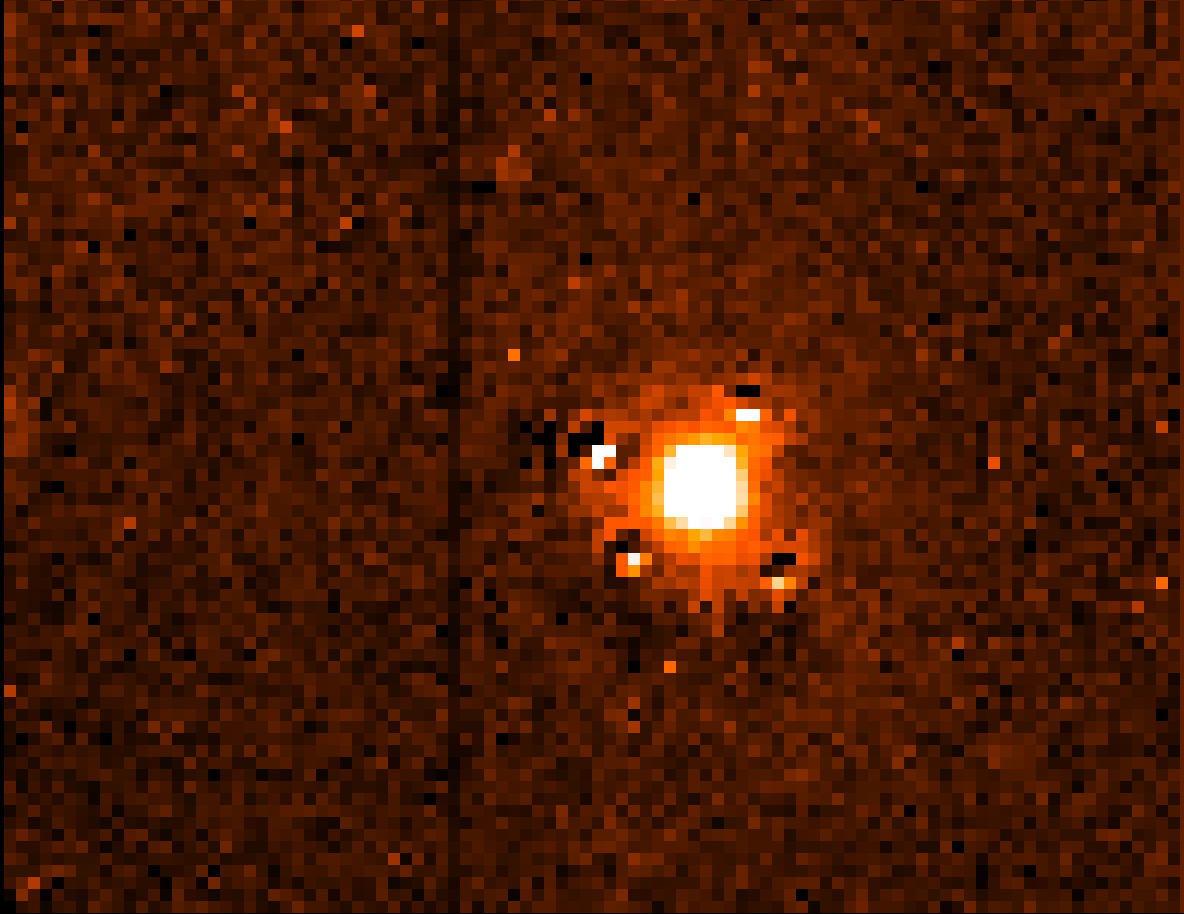} \\[0.25cm]
\end{tabular}
\label{fig_sub}
\end{figure*}

\indent Although it is valid to the first order, the hypothesis of symmetry around $r_{\rm{max}}$ may not always apply perfectly. Moreover, our symmetry criterion is more stable when restricted closer to the arc maximum, where the signal clearly dominates the noise. To extend the arc estimate to lower intensities, we fit a de Vaucouleurs law on each of the arc wings. We then compute a linear combination (per wing) of the fitted values and the datapoints, with weights so that the datapoints dominate around the centre of symmetry and the fitted values dominate closer to the wings (see Fig. \ref{fig_wings}). That way we obtain a numerical profile that is a close approximation of the true arc radial light distribution.\\\\
\indent Finally, for each trace within the sector, we scale that numerical profile to better fit the radial intensities of each angular coordinate. We repeat the process on each sector, yielding for each one a numerical radial profile of the arc. We end up with as many arc profiles as traces, each corresponding to a single angular coordinate. We thus have a map of the arc intensity as a function of galactocentric coordinates $(r, \theta)$. To reconstruct a 2-D image of the arc in cartesian coordinates, we compute the intensity of each pixel by interpolating on its direct neighbours on that map. Finally, we can subtract that 2-D image from the data frame. The arc subtraction has been performed on each individual image of four out of seven galaxies in our sample. Indeed, on the remaining three systems, there was no visible diffuse lensed signal contaminating the galaxy light, no significant "bump" in their radial profile. The results are shown in Fig. \ref{fig_sub}. Once all the lensed signal, both point-like and diffuse, has been subtracted from the data frames, we can proceed to the shape parameters measurement, as explained in Sect \ref{sec_shape}.

\subsection{Measuring the shape parameters}\label{sec_shape}

\indent The second aim of this work is to compute, with the highest possible accuracy, the shape parameters of the lenses (the PA, $\varepsilon$, and $r_{\rm{eff}}$). Each parameter has been calculated individually, in order to avoid the problem of potential local minima in the parameters space. The methods are described in great details in \citetalias{biernauxetal2016}. Since the publication of that paper, we have added the arc subtraction to the pre-processing of our sample. Therefore, we conducted those measurements again. The results remain unchanged for the position angle and the ellipticity. An extensive description of those parameters' measurements, can be found in \citetalias{biernauxetal2016} and in the associated table summarising our results on the CDS website \citep{B16CDS}. However, we have slightly modified the $r_{\rm{eff}}$ measurement method, improved its values, and computed a more secure error bar. \\

\indent In \citetalias{biernauxetal2016}, the LRM was based on determining the slope of a linear relationship between $\ln{I}$ and $r^{1/4}$, defined by the logarithm of Eq.  \ref{eq_dVc};
\begin{center}
\begin{equation}
\centering
\ln{I} = \ln{I_{\rm{0}}} -  k\left(\frac{r}{r_{\rm{eff}}}\right)^{1/4}
\label{eq_lndVc}
\end{equation}
\end{center}
assuming a de Vaucouleurs brightness distribution. The radial profile of each lens has been obtained by computing its isophotes, using one-pixel-wide masks of increasing semi-minor axis. Then, a de Vaucouleurs profile convolved by the instrument PSF was generated and its shape parameters were fine-tuned until the slope of this synthetic lens in the ($\ln{I}$~;~$r^{1/4}$) space matches the measured slope on the data frame. In this work, we have chosen to proceed in a similar way, only we do not use the slope of $\ln{I}(r^{1/4})$ as a validity criterion for the model. Instead, we adjust the shape parameters until they minimise the $\chi^{2}$ between the datapoints and the convolved model radial profile in the ($\ln{I}$~;~$r^{1/4}$) space. Because we do not have any information about $I_{\rm{0}}$ at that point, for each try, the convolved model logarithmic profile is translated by a constant ($\Delta$) that is an average of the difference between matching datapoints from the convolved model and the data frame radial profiles. This process is less time-consuming than the LRM and we find it to be more trustworthy as it grants less importance to the outliers. \\

\begin{figure*}[!pht]
%\captionlistentry[table]{A table beside a figure}
%\captionsetup{labelformat=andtable}
%\caption{}
\begin{minipage}[c]{0.46\linewidth}
\centering
\captionof{table}{Value of $r_{\rm{eff}}$ from \citetalias{biernauxetal2016} and after update in this work.}
\begin{tabular}{c c c c }
\hline
\textit{Index} & \textit{System} & \textit{\citetalias{biernauxetal2016}  $r_{\rm{eff}}$ ('')} & \textit{\textit{$r_{\rm{eff}}$ ('')}} \\ 
\hline
 & & & \\ 
        1 & MG0414+0534 & $ 0.737 \pm 0.096 $ & $0.660 \pm 0.100$ \\
        2 & HE0435-1223 & $ 0.901 \pm 0.071 $ & $0.872 \pm 0.076$ \\
        3 & RXJ0911+0551 &  $ 0.878 \pm 0.187 $ & $0.869 \pm 0.207$ \\
        4 & SDSS0924+0219 & $ 0.295 \pm 0.044 $ & $0.253 \pm 0.062$ \\
        5 & PG1115+080 & $ 0.433\pm  0.086 $ & $0.443 \pm 0.092$ \\
        6 & SDSS1138+0314 & $ 0.352 \pm 0.043 $ & $0.199 \pm 0.085$ \\
        7 & B1422+231 & $ 0.114 \pm 0.059 $& $0.107 \pm 0.056$ \\
& & & \\
\hline
\end{tabular}
\label{tab_reff}
\end{minipage}
\hfill
\begin{minipage}[c]{0.46\linewidth}
\centering
\captionof{figure}{New $r_{\rm{eff}}$ values as a function of \citetalias{biernauxetal2016} values, plotted with their error bars, for a quicker visualization. The solid line represents the equality between values from \citetalias{biernauxetal2016} and this work.}
\includegraphics[scale=0.45]{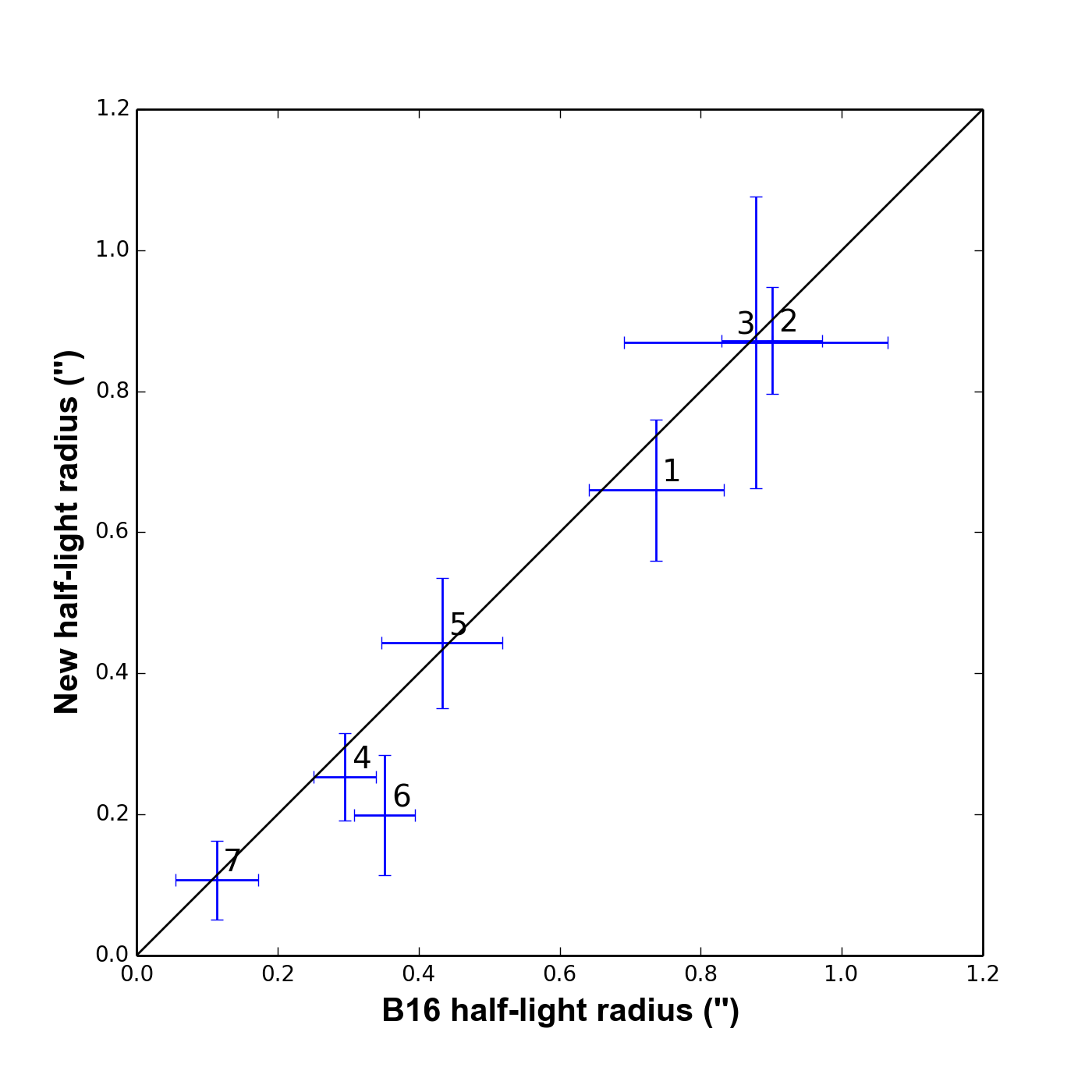}
\label{fig_reff}
\end{minipage}
\end{figure*}

\begin{figure*}[!pht]
\caption{Left to right: data frame after arc and point-source subtraction, best model and residuals for each system. The circle depicts the region of interest for the measurement of $r_{\rm{eff}}$. The residual maps correspond to a $\pm 3\sigma$ scale, white corresponding to $> 3\sigma$ and black, $< 3\sigma$. Only one data frame is shown for each system.}
%\textit{From top to bottom}: MG0414+0534, HE0435-1223, RXJ0911+0551, SDSS0924+0219, PG1115+080, SDSS1138+0314, B1422+231. 
\centering
\begin{tabular}{c c c}
 & & \\
\includegraphics[scale=0.1265]{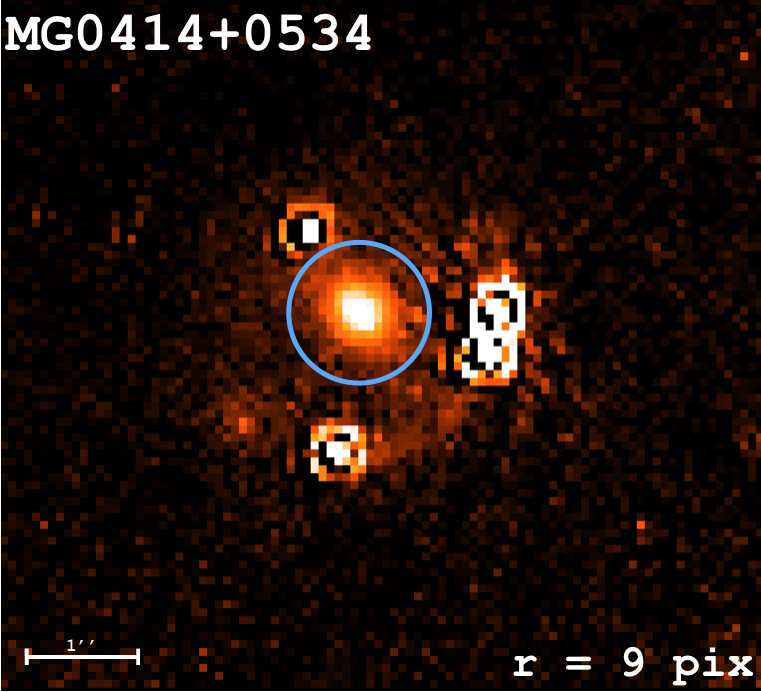} & \includegraphics[scale=0.1]{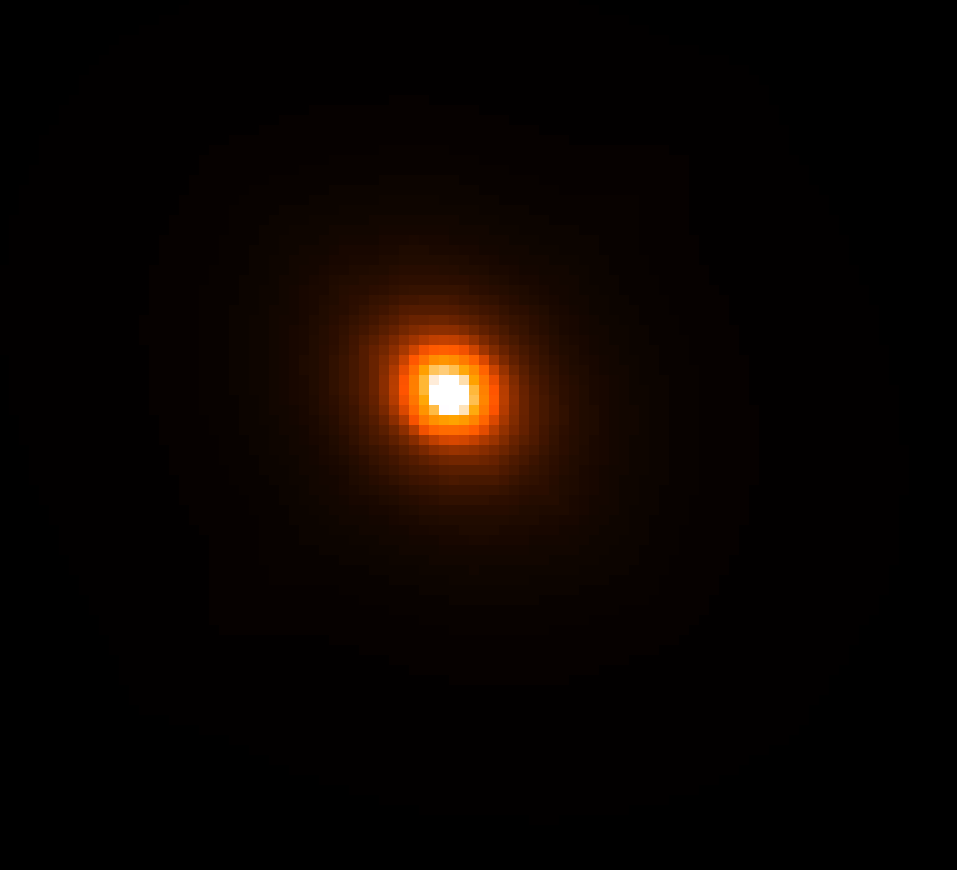} & \includegraphics[scale=0.1125]{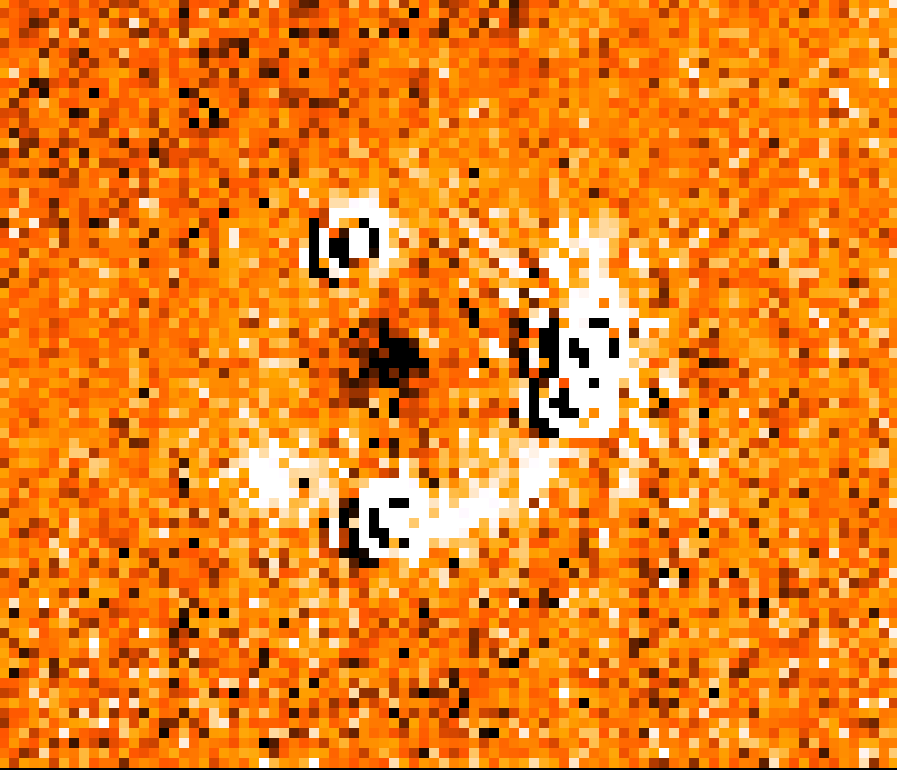} \\[0.25cm]
\includegraphics[scale=0.1265]{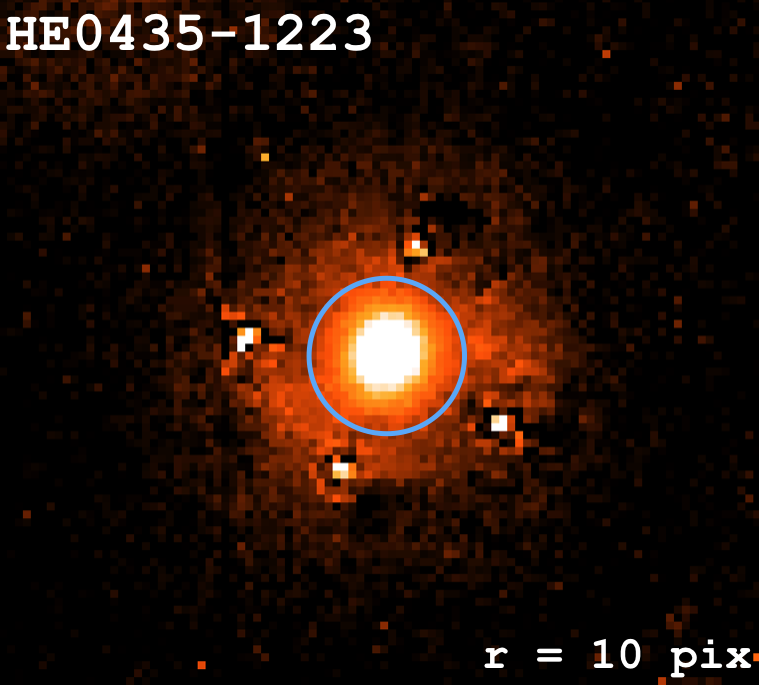} & \includegraphics[scale=0.1]{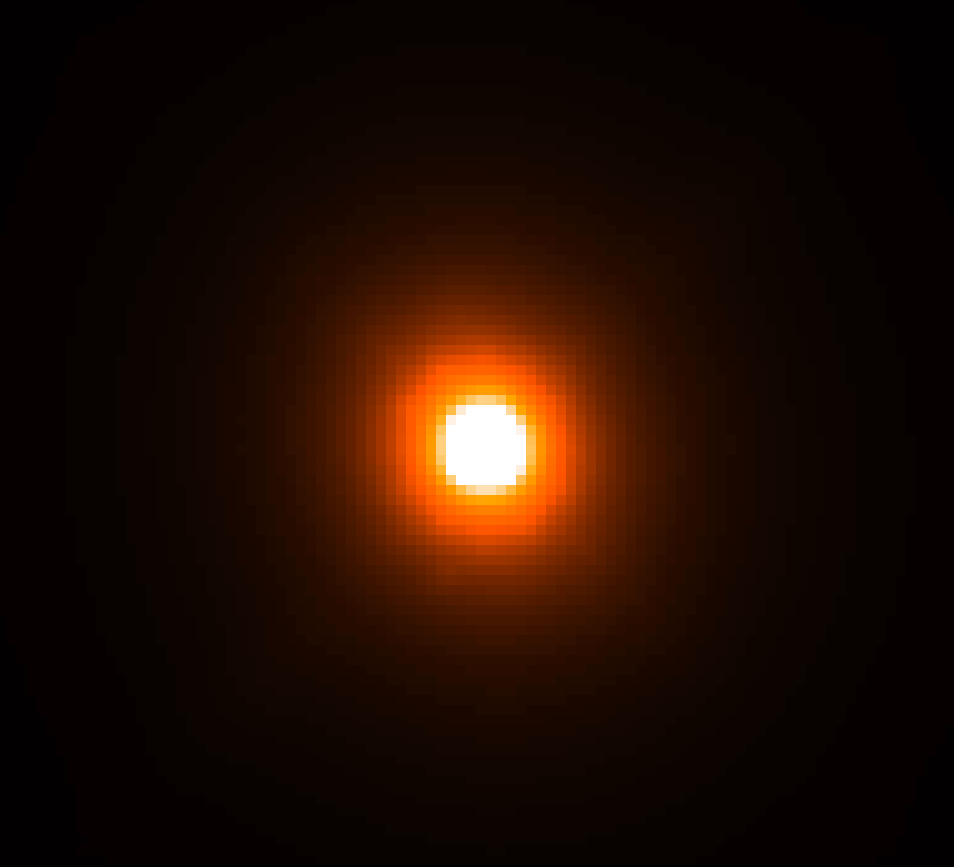} & \includegraphics[scale=0.1125]{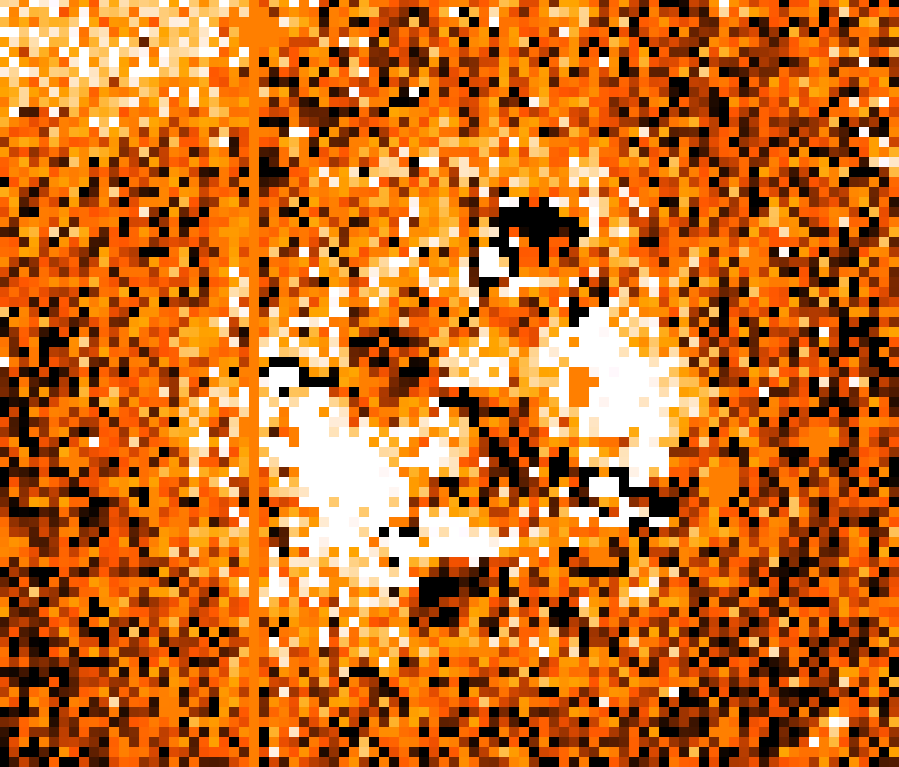} \\[0.25cm]
\includegraphics[scale=0.1265]{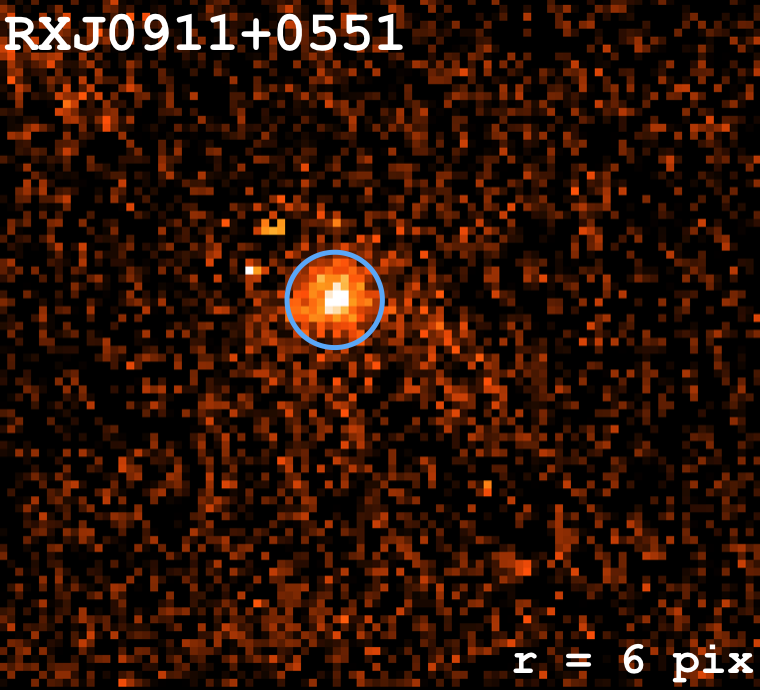} & \includegraphics[scale=0.1]{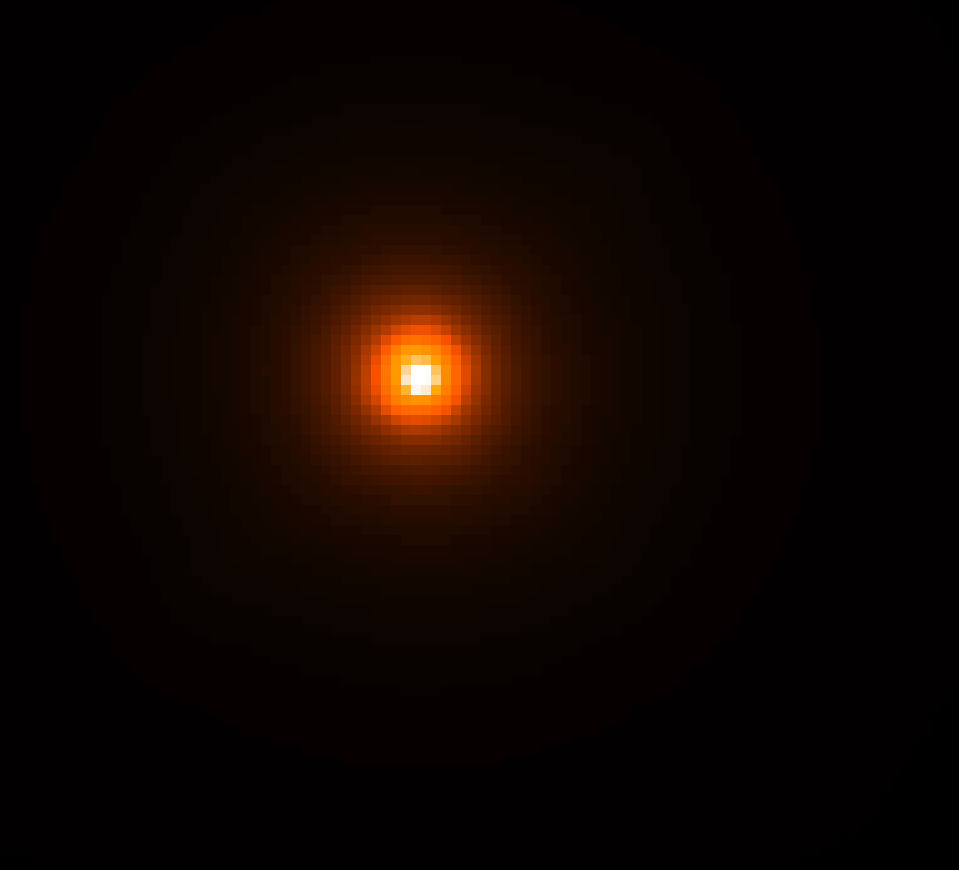} & \includegraphics[scale=0.1125]{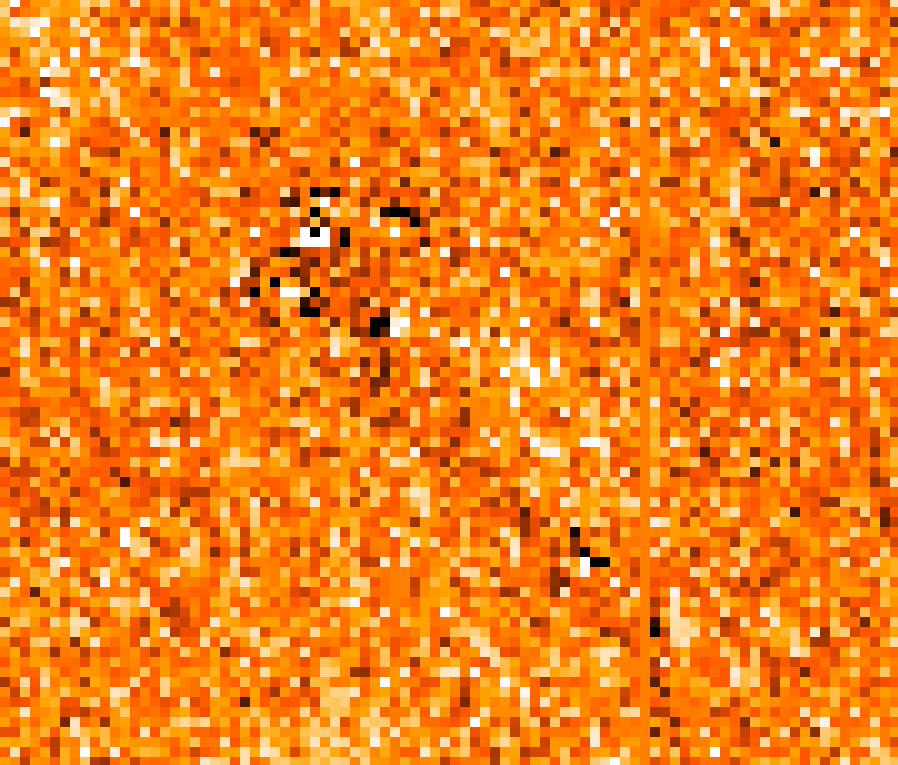} \\[0.25cm]
\includegraphics[scale=0.1265]{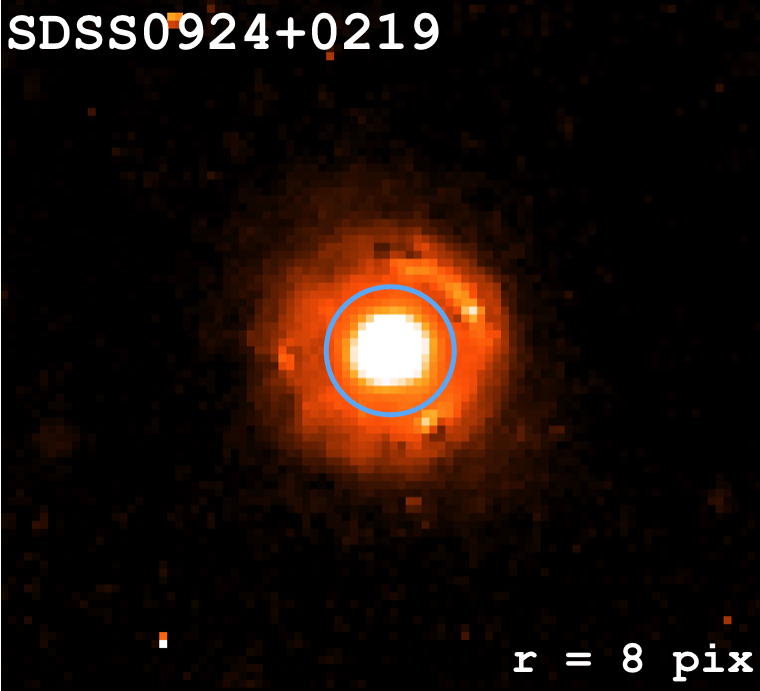} & \includegraphics[scale=0.1]{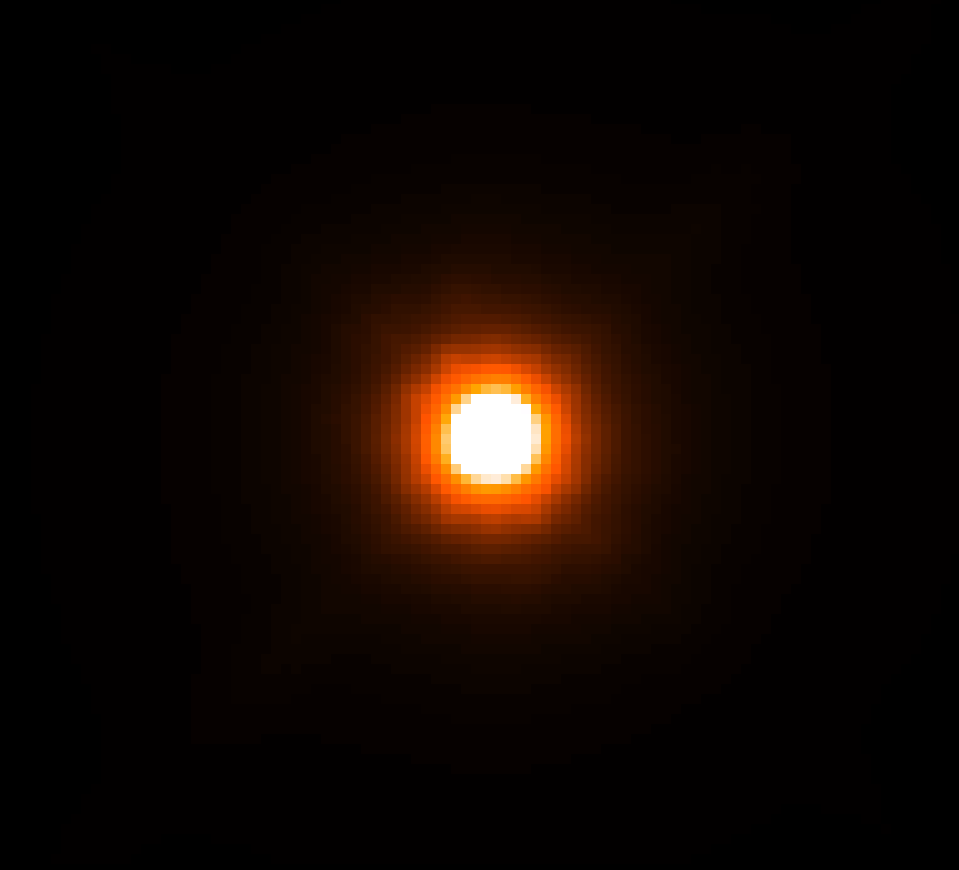} & \includegraphics[scale=0.1125]{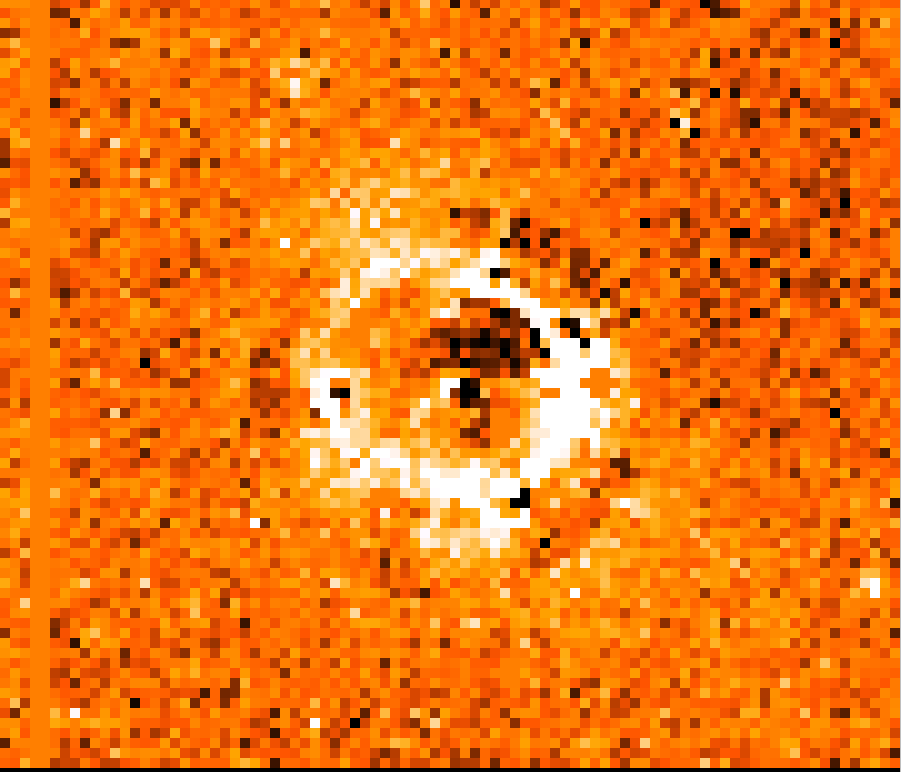} \\[0.25cm]
\includegraphics[scale=0.1265]{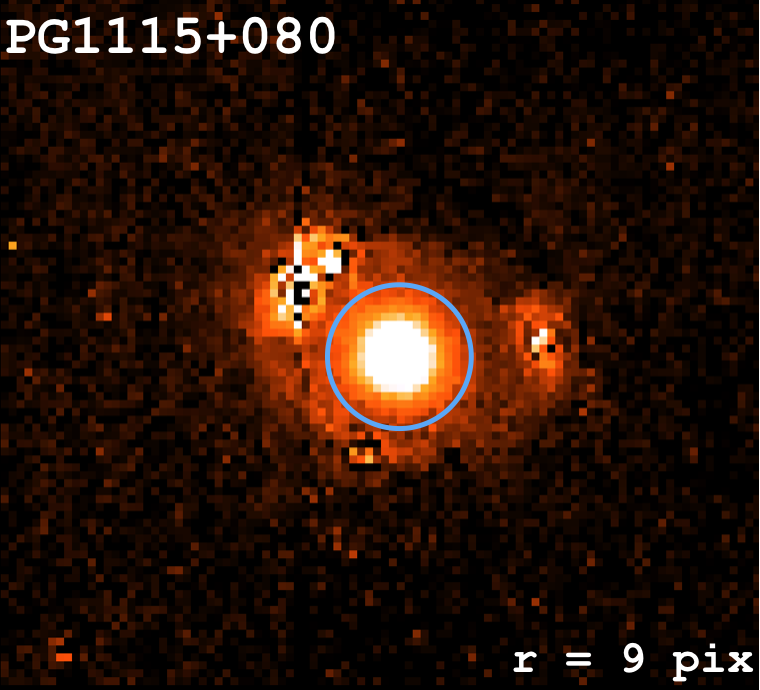} & \includegraphics[scale=0.1]{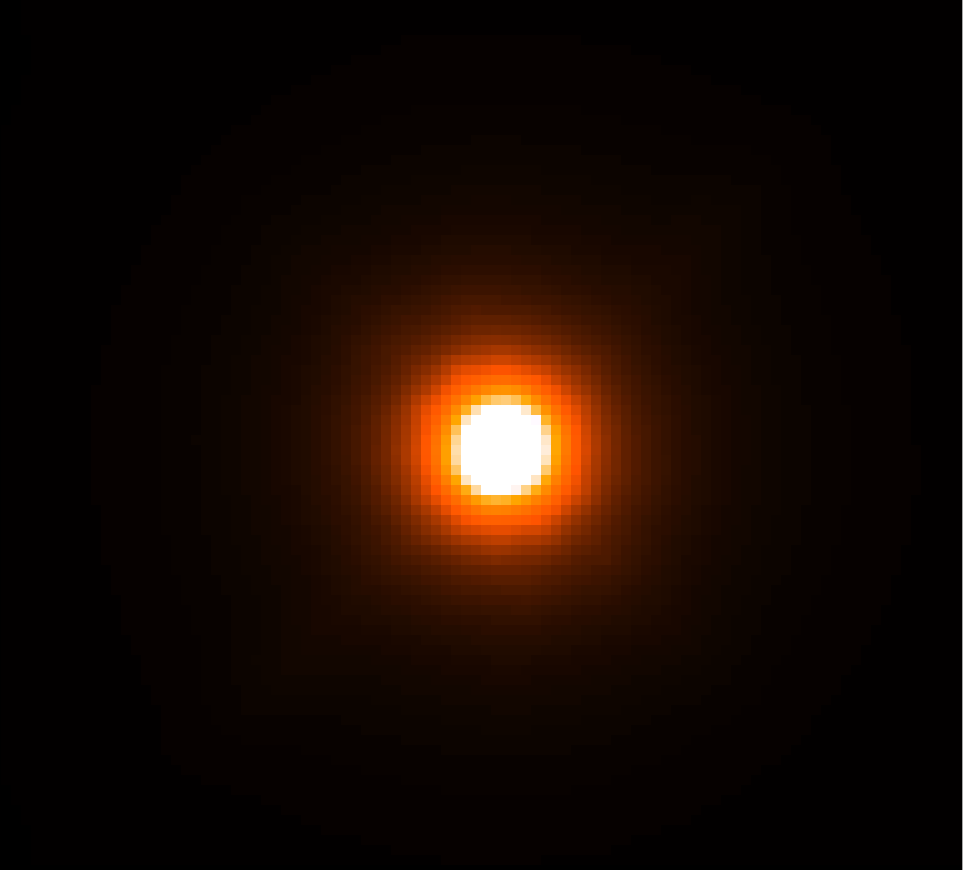} & \includegraphics[scale=0.1125]{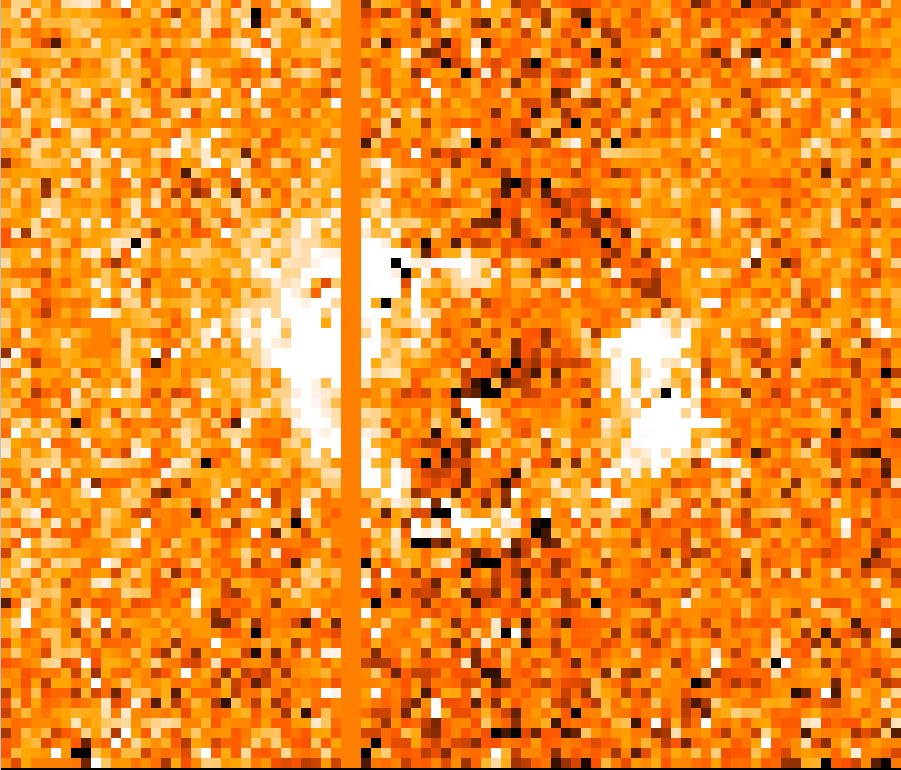} \\[0.25cm]
\includegraphics[scale=0.1265]{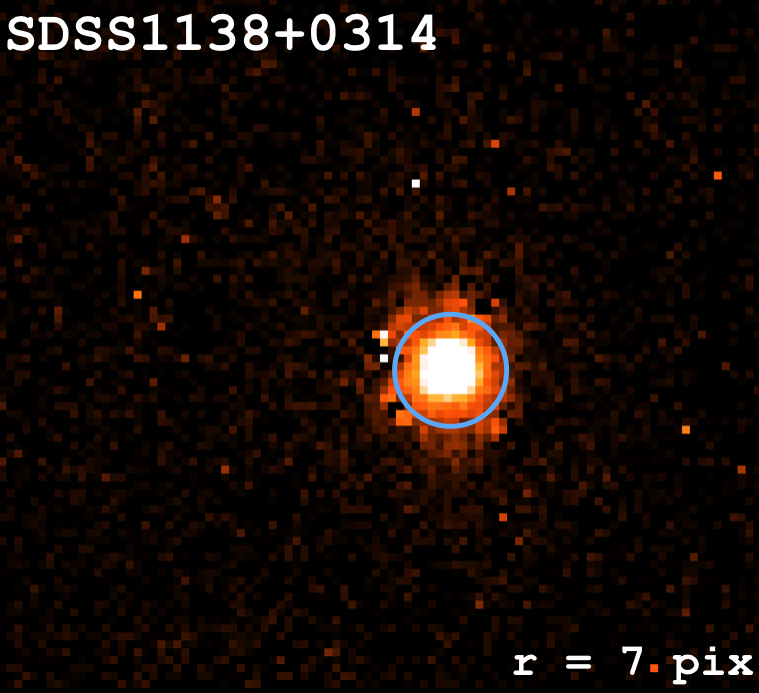} & \includegraphics[scale=0.1]{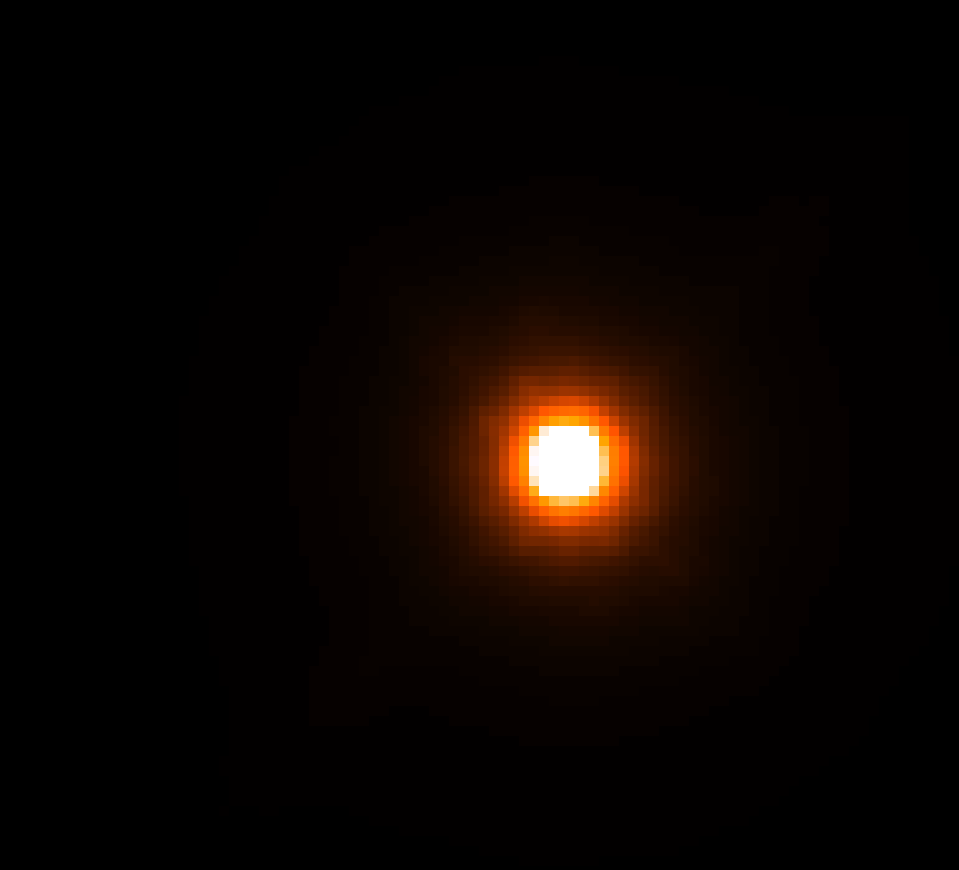} & \includegraphics[scale=0.1125]{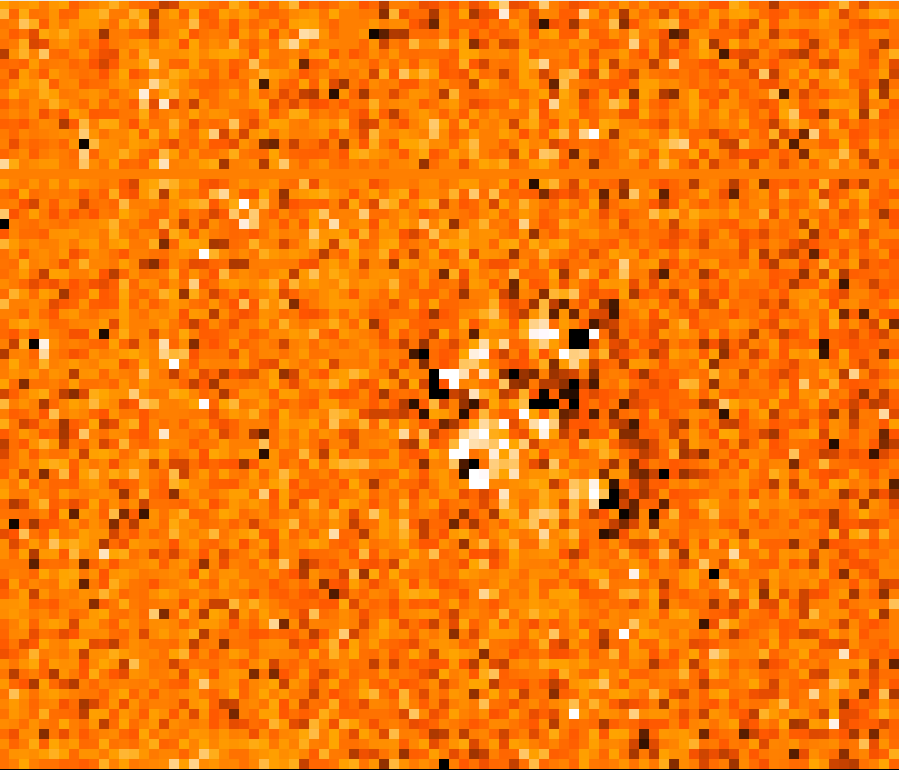} \\[0.25cm]
\includegraphics[scale=0.1265]{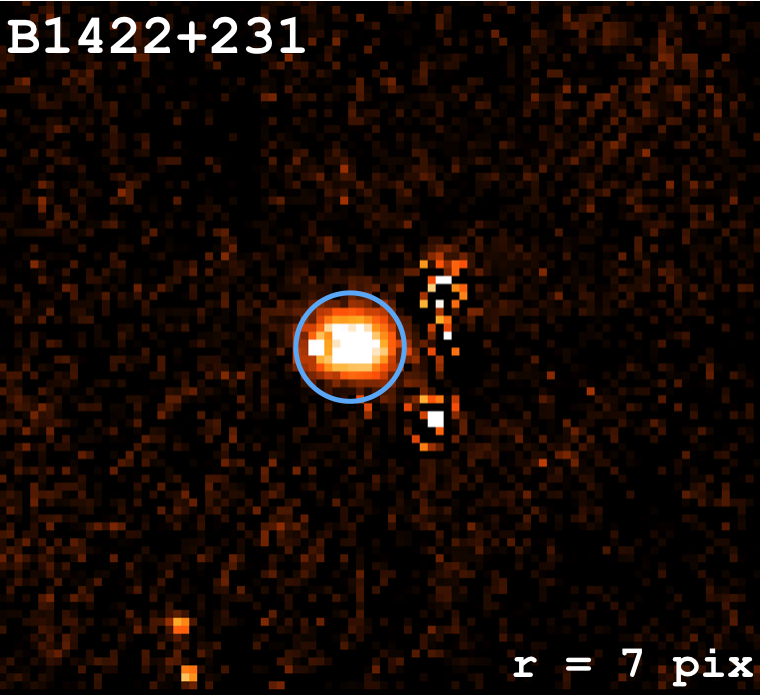} & \includegraphics[scale=0.1]{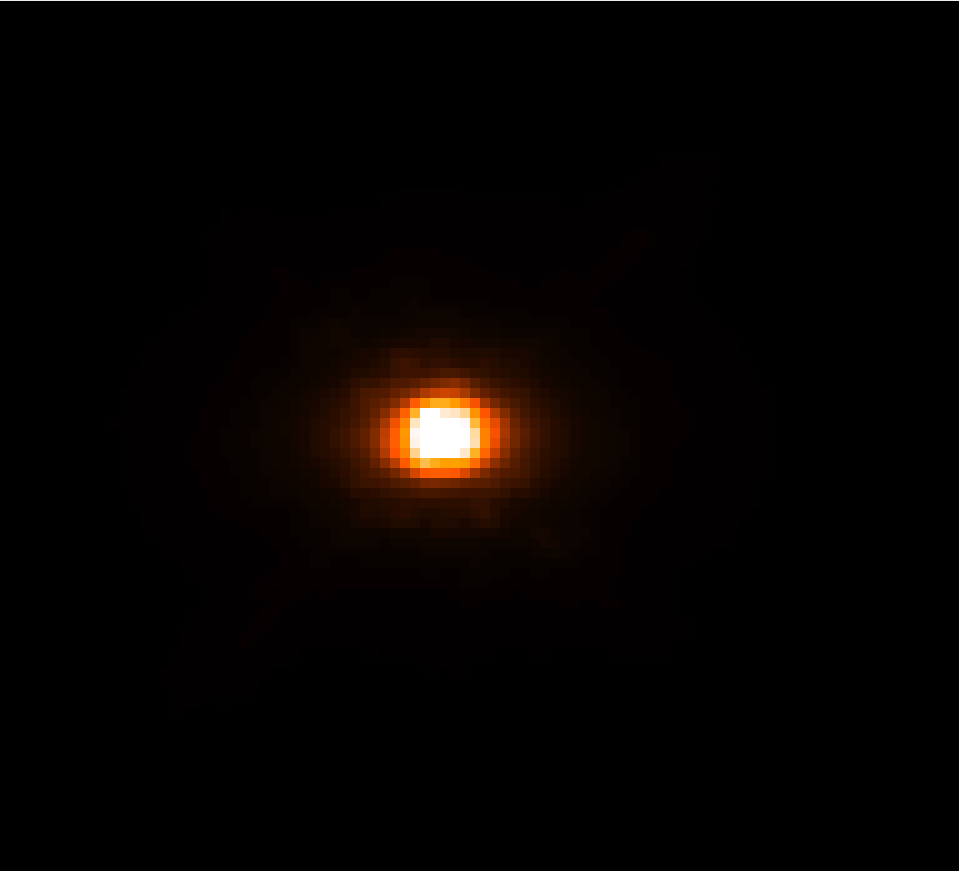} & \includegraphics[scale=0.1125]{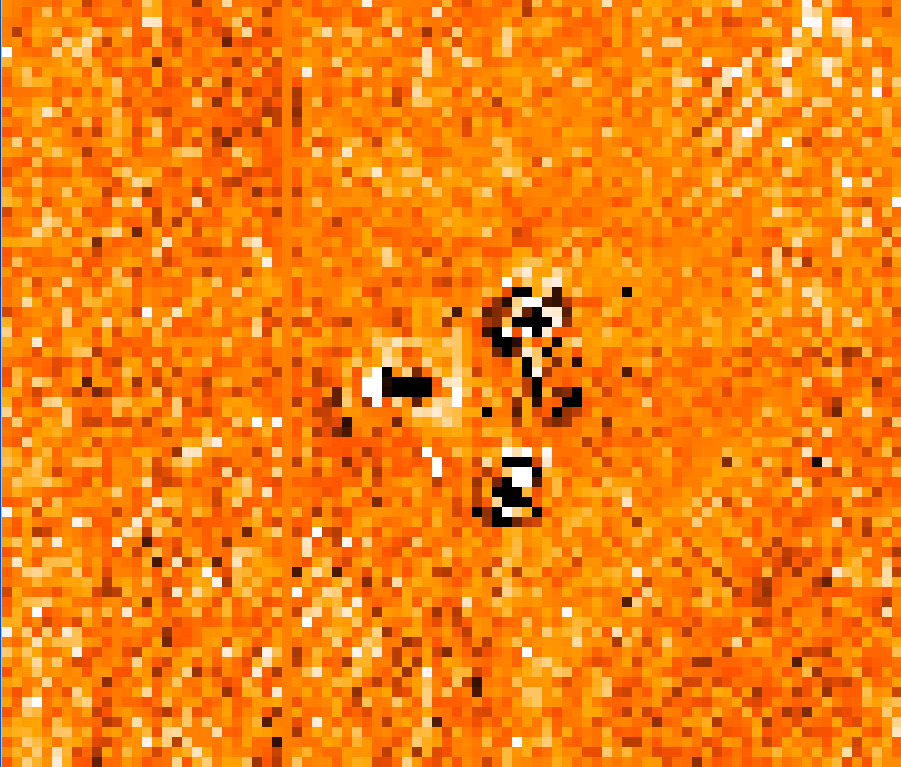} \\[0.25cm]
\end{tabular}
\label{fig_resi}
\end{figure*}

\indent In conclusion, even though the arc intensity is not dominant in the portion of the galaxy that is taken into account for the $r_{\rm{eff}}$ measurement, it still produces some contamination. Thanks to the arc subtraction, that contamination is removed and more points in the outer parts of the lens galaxy can be taken into account for the computation of its radial profile, which improves the $r_{\rm{eff}}$ measurement, as discussed in Sect \ref{sec_results}. The region of interest for this new LRM is shown in Fig. \ref{fig_resi}. As in \citetalias{biernauxetal2016}, its radius is determined visually by the user. The $r_{\rm{eff}}$ values obtained in \citetalias{biernauxetal2016} have been updated, as listed in Table \ref{tab_reff}.
\subsection{Computing the integrated luminosity within the Einstein radius}\label{sec_integ}

\indent The final aim of this work is to calculate the integrated flux in the H-band within the Einstein radii $r_{\rm{E}}$ of our seven lenses. We used the $r_{\rm{E}}$ values from \citet{Sluseetal2012a}, more specifically, from their Table 7. We list them in Table \ref{tab_k}.

Similarly to what is performed for the shape parameters, we aim at a simple, 1-D measurement method for the central intensity parameter $I_{\rm{0}}$; we scan a wide interval of $I_{\rm{0}}$ values and refine it until it minimises the $\Delta$ translation constant between the model and the data radial profiles. Once in possession of the correct $I_{\rm{0}}$ value, we can produce an image of the deconvolved galaxy profile, which is thus the best analytical model of the lensing galaxy brightness distribution. This is the image that we use to compute the integrated flux within the Einstein radius $F_{\rm{H}}$. The measurement is simply performed by integrating the total signal in an aperture centred on the lens of a radius equal to $r_{\rm{E}}$, then converting this value into an H-band flux, taking into account the instrument gain, exposure time, and a K-correction. The latter corrects from the offset between the redshifted lens spectrum and the NICMOS instrument bandpass, and is computed based on an elliptical galaxy template spectrum synthesised using the P\'egase software \citep{Pegase}. The K-correction results as a factor K such that $F_{\rm{H}} = KF_{\rm{\lambda}}$, where $F_{\rm{\lambda}}$ is the measured flux and $F_{\rm{H}}$ is the corrected flux. All three values are given for each system in Table \ref{tab_k}. We also give $F_{\rm{H}}$ expressed in solar units, using a value of $F_{\rm{H,\odot}}$ \citep{Colinaetal1996} at the distance of each lens for a WMAP \citep{Lewis2008} cosmological model, $\Omega_{\Lambda} = 0.73; \;\Omega_{M} = 0.27; \;h = 0.71$. This set of cosmological parameters has been chosen for the purpose of a comparison with results from \citetalias{biernauxetal2016}, \citet{Chantryetal2010}, and \citet{Sluseetal2012a}.\\
\begin{table*}[!pht]
\caption{Integrated flux within an aperture of radius $r_{\rm{E}}$ before ($F_{\rm{\lambda}}$) and after ($F_{\rm{H}}$) K-correction.}
\centering
\begin{tabular}{c c c c c c }\hline
\textit{System } & $r_{\rm{E}}$ ('') (1) &  $F_{\rm{\lambda}} (\rm{erg\;s^{-1}\;cm^{-2}}$ \AA$^{-1} )$ & \textit{K-correction} & $F_{\rm{H} }(\rm{erg\;s^{-1}\;cm^{-2}}$ \AA$^{-1} )$ & $F_{\rm{H}}(10^{11} F_{\rm{H,\odot}})$ \\ \hline
 & & & & & \\
MG0414+0534 & 1.182 &  $8.329\;10^{-18}$ & $0.547$ & $4.556\;10^{-18}$ & $3.505 \pm 0.839$ \\
HE0435-1223 & 1.201 & $1.682\;10^{-17}$ & $0.674$ & $1.134\;10^{-17}$ & $1.391 \pm 0.203$\\
RXJ0911+0551 & 1.120 & $7.211\;10^{-18}$ & $0.584$ & $4.212\;10^{-18}$ &$1.872 \pm 0.044$\\
SDSS0924+0219 & 0.874 & $1.468\;10^{-17}$ & $0.699$ & $1.026\;10^{-17}$ &$0.877 \pm 0.173$\\
PG1115+080 & 1.145 & $2.080\;10^{-17}$& $0.739$ & $1.537\;10^{-17}$ & $0.761 \pm 0.091$\\
SDSS1138+0314 & 0.664 & $7.919\;10^{-18}$ & $0.676$ &$5.353\;10^{-18}$  & $0.641 \pm 0.075$\\
B1422+231 & 0.771 & $7.380\;10^{-18}$ & $0.725$ &$5.350\;10^{-18}$ & $0.322 \pm 0.094$\\
 & & & & & \\\hline
\end{tabular}
\label{tab_k}
\tablebib{(1) \citet{Sluseetal2012a}.}
\end{table*}  

%%%%%%%%%%%%%%%%%%%%%%%%%%%%%%%%%%%%%%%%%%%%%%
\section{Error calculation}\label{sec_err}
\subsection{On the shape parameters}

\indent We have conducted a thorough error calculation on the shape parameters taking into account five sources of systematic errors: 
the determination of the $x-$ and $y-$positions of the deflected images, of their intensities, of the galaxy centre coordinates, of the sky background, and arc subtraction. The process for the first four items is detailed in \citetalias{biernauxetal2016}: we compute a dispersion error on each of these systematics values among all the data frames of each system, then conduct the shape parameters measurement again with the values shifted by an offset of the same magnitude as their error bar, one at a time. In this section, we focus on the new source of systematic errors, that is the subtraction of the arc. We used a similar approach that is based on building a $1\sigma$ error image of the arc and then measuring the shape parameters after subtracting the sum of the arc image and its $1\sigma$ image.\\

\indent Each pixel of the data frame is affected by an uncertainty on its intensity, that is an output of the HST-NICMOS pipeline data reduction.\footnote{Or that can be inferred from the data quality map and the photon noise.} We analytically compute the propagation of these uncertainties at each step of the arc reconstruction: the stacking and collapsing of traces within each sector, the fitting of a de Vaucouleurs law with the residuals symmetry criterion, the linear combination between this fit and arc datapoints, the re-scaling on individual traces, and the 2-D reconstruction of the arc. This yields a 1$\sigma$ pixel map for the arc. To evaluate the error from the arc subtraction on the measurement of $r_{\rm{eff}}$, for example, we conduct it twice, once after a subtraction of the correct arc image, and once deliberately subtracting too much arc, adding the 1$\sigma$ value for each pixel in the arc image. The magnitudes of these error bars on $r_{\rm{eff}}$, as well as for each systematic error source, are listed in Table \ref{tab_errreff}. We find that the arc subtraction did not cause any noticeable error on the measurement of the PA nor of $\varepsilon$. Therefore, we do not update those parameters' error calculations since \citetalias{biernauxetal2016} and their error budget can be found in the \citetalias{biernauxetal2016} CDS table \citep{B16CDS}. \\

\begin{table*}[pht]
\caption{Error budget on $r_{\rm{eff}}$. From left to right: dispersion, sky subtraction, positions and intensities of the sources, position of the galactic centre, and subtraction of the arc. N.A. stands for "not applicable".}
\centering
\begin{tabu}{c c c  c  c  c  c  c  c  c c }
\hline
\textit{System} & $r_{\rm{eff}}$ \textit{('')} & $\sigma_{\rm{rand}}$ & $\sigma_{\rm{sky}}$ & $\sigma_{\rm{xs}}$ & $\sigma_{\rm{ys}}$ & $\sigma_{\rm{Is}}$ & $\sigma_{\rm{xg}}$ & $\sigma_{\rm{yg}}$ & $\sigma_{\rm{arc}}$ & \textit{Total error} \\ 
\hline
& & & & & & & & & &\\
MG0414+0534 & 0.660 & 0.044 & 0.010 & 0.029 & 0.029 & 0.028 & 0.017 & 0.073 & N.A. & 0.100  \\
HE0435-1223 & 0.872 & 0.008 & 0.001 & 0.026 & 0.027 & 0.034 & 0.030 & 0.034 & 0.034 & 0.076 \\
RXJ0911+0551 & 0.869 & 0.106 & 0.018 & 0.105 & 0.102 & 0.063 & 0.077 & 0.007 & N.A. & 0.207 \\
SDSS0924+0219 & 0.253 & 0.012 & 0.031 & 0.009 & 0.009 & 0.009 & 0.007 & 0.025 & 0.043 & 0.062  \\
PG1115+080 & 0.443 & 0.012 & 0.041 & 0.032 & 0.032 & 0.032 & 0.017 & 0.047 & 0.033 & 0.092 \\
SDSS1138+0314 & 0.199 & 0.005 & 0.000 & 0.002 & 0.002 & 0.005 & 0.008 & 0.039 & 0.075 & 0.085 \\
B1422+231 & 0.107 & 0.012 & 0.014 & 0.030 & 0.026 & 0.027 & 0.015 & 0.014 & N.A. & 0.056\\
& & & & & & & & & & \\
\hline
\end{tabu}
\label{tab_errreff}
\end{table*}   

\subsection{On the integrated luminosity}

\indent We adopt a similar procedure to compute an error bar on $F_{\rm{H}}$. We consider the same sources of systematic errors (sky subtraction, point sources positions and intensities, and arc subtraction), conduct the $F_{\rm{H}}$ measurement with their values shifted by an offset equal to their $1\sigma$ error bar, and get their contribution to the total $F_{\rm{H}}$ error budget. On top of that, we take into account the effect from a miscalculation of $r_{\rm{eff}}$ on the $F_{\rm{H}}$ measurement. Indeed, if $r_{\rm{eff}}$ is overestimated, for example, the central intensity $I_{\rm{0}}$ would be underestimated and thus give a wrong result for the integrated flux. To quantify this effect, we conduct the $F_{\rm{H}}$ measurement (from the $I_{\rm{0}}$ determination) with a set of parameters that include an overestimate value of $r_{\rm{eff}} + \sigma_{\rm{r_{\rm{eff}}}}$. Because we already compute an error bar on $F_{\rm{H}}$ from the point source positions and intensities and from the sky subtraction, the $\sigma_{\rm{r_{\rm{eff}}}}$ value we use here only results from the combination of its dispersion and its error bar from the galaxy centre position ($\sigma_{\rm{rand}}$, $\sigma_{\rm{xg}}$ and $\sigma_{\rm{yg}}$). We find that while $I_{\rm{0}}$ and $r_{\rm{eff}}$ are tightly correlated, the errors on the PA and $\varepsilon$ did not influence the $F_{\rm{H}}$ measurement. The error budget of $F_{\rm{H}}$ is given in Table \ref{tab_errflux}.

\begin{table*}[pht]
\caption{Error budget on $F_{\rm{H}}$. From left to right: dispersion, sky subtraction, positions and intensities of the sources, measurement of $r_{\rm{eff}}$, and subtraction of the arc. N.A. stands for "not applicable".}
\centering
\begin{tabu}{ c c c  c  c  c  c  c  c  c }
\hline
\textit{System} & $F_{\rm{H}}  (10^{11} F_{\rm{H}, \odot}$) & $\sigma_{\rm{rand}}$ & $\sigma_{\rm{sky}}$ & $\sigma_{\rm{xs}}$ & $\sigma_{\rm{ys}}$ & $\sigma_{\rm{Is}}$ & $\sigma_{\rm{reff}}$ & $\sigma_{\rm{arc}}$ & \textit{Total error} \\ 
\hline
& & & & & & & & &\\
MG0414+0534 & 3.505 & 0.823 & 0.340 & 0.398 & 0.375 & 0.379 &  0.372 & N.A. & 0.839  \\
HE0435-1223 & 1.391 & 0.029 & 0.092 & 0.080 & 0.080 & 0.079 & 0.071 & 0.088 & 0.203 \\
RXJ0911+0551 & 1.872 & 0.024 &  0.030 &  0.001 & 0.009 & 0.007 &  0.017 & N.A. & 0.044 \\
SDSS0924+0219 & 0.877 & 0.005 & 0.074 & 0.075 & 0.077 & 0.077 &  0.072 & 0.043 & 0.173  \\
PG1115+080 & 0.761 & 0.004 & 0.032 &  0.035 & 0.035  & 0.032 & 0.035 & 0.051 & 0.091 \\
SDSS1138+0314 & 0.641 & 0.011 & 0.018  & 0.022 & 0.023 & 0.023 & 0.025 & 0.055 & 0.075 \\
B1422+231 & 0.322 & 0.006 & 0.041 & 0.043 & 0.038 & 0.045 & 0.042 & N.A. & 0.094 \\
& & & & & & & & & \\
\hline
\end{tabu}
\label{tab_errflux}
\end{table*}

%%%%%%%%%%%%%%%%%%%%%%%%%%%%%%%%%%%%%%%%%%%%%%
\section{Results and discussion}\label{sec_results}

\indent The results of the half-light radii measurements are presented in Table \ref{tab_reff} and compared to the \citetalias{biernauxetal2016} results. The new $r_{\rm{eff}}$ are on average $11 \pm 6$ \% smaller than the \citetalias{biernauxetal2016} values, but their error bars overlap, except for SDSS1138+0314. This decrease can be expected as we subtract the contribution of the arc that leads to a positive bias on $r_{\rm{eff}}$. Thanks to the subtraction of the arc, more datapoints can be taken into account for the $r_{\rm{eff}}$ measurement; we can consider a portion of the galaxy with a radius that is one or two pixels, or 17\% on average, larger than before. Although this may seem like a small increase, it is of considerable importance because the $r_{\rm{eff}}$ value is highly sensitive to the wings of the de Vaucouleurs profile. Moreover, because the convolution mostly affects the central regions of the galaxy, we choose to ignore the first couple of pixels from the centre out, hence the interest in reaching as far out from the galactic centre as possible. The $r_{\rm{eff}}$ values obtained for the systems where no arc was subtracted have also changed, because of the new LRM. Changing the validity criterion in the LRM from only the slope of the model to the $\chi^{2}$ takes into account the individual uncertainties of the datapoints, granting less importance to the outliers and giving a more trustworthy $r_{\rm{eff}}$ value. Both these changes with regards to \citetalias{biernauxetal2016} constitute an improvement of the $r_{\rm{eff}}$ values.\\

\begin{figure*}[!pht]
\caption{Plots of the residual curvature when the $\ln I$ vs $r^{1/4}$ radial profile of the convolved best de Vaucouleurs model has been subtracted from that of the data frame. The stars show the residual curvature from the \citetalias{biernauxetal2016} models, the crosses, from this work.}
%\textit{From top to bottom}: MG0414+0534, HE0435-1223, RXJ0911+0551, SDSS0924+0219, PG1115+080, SDSS1138+0314, B1422+231. 
\centering
\begin{tabular}{c c }
\includegraphics[scale=0.45]{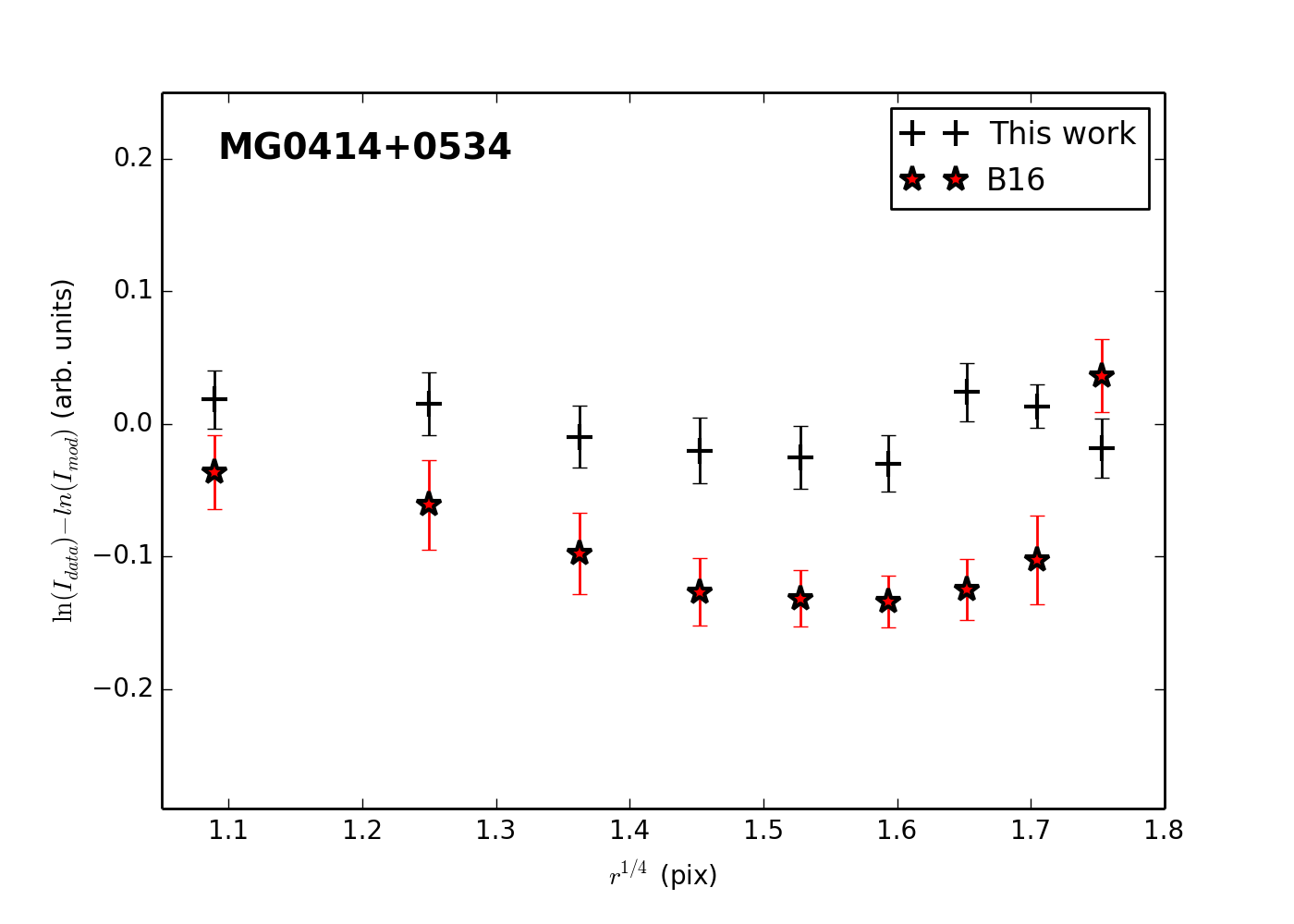} & \includegraphics[scale=0.45]{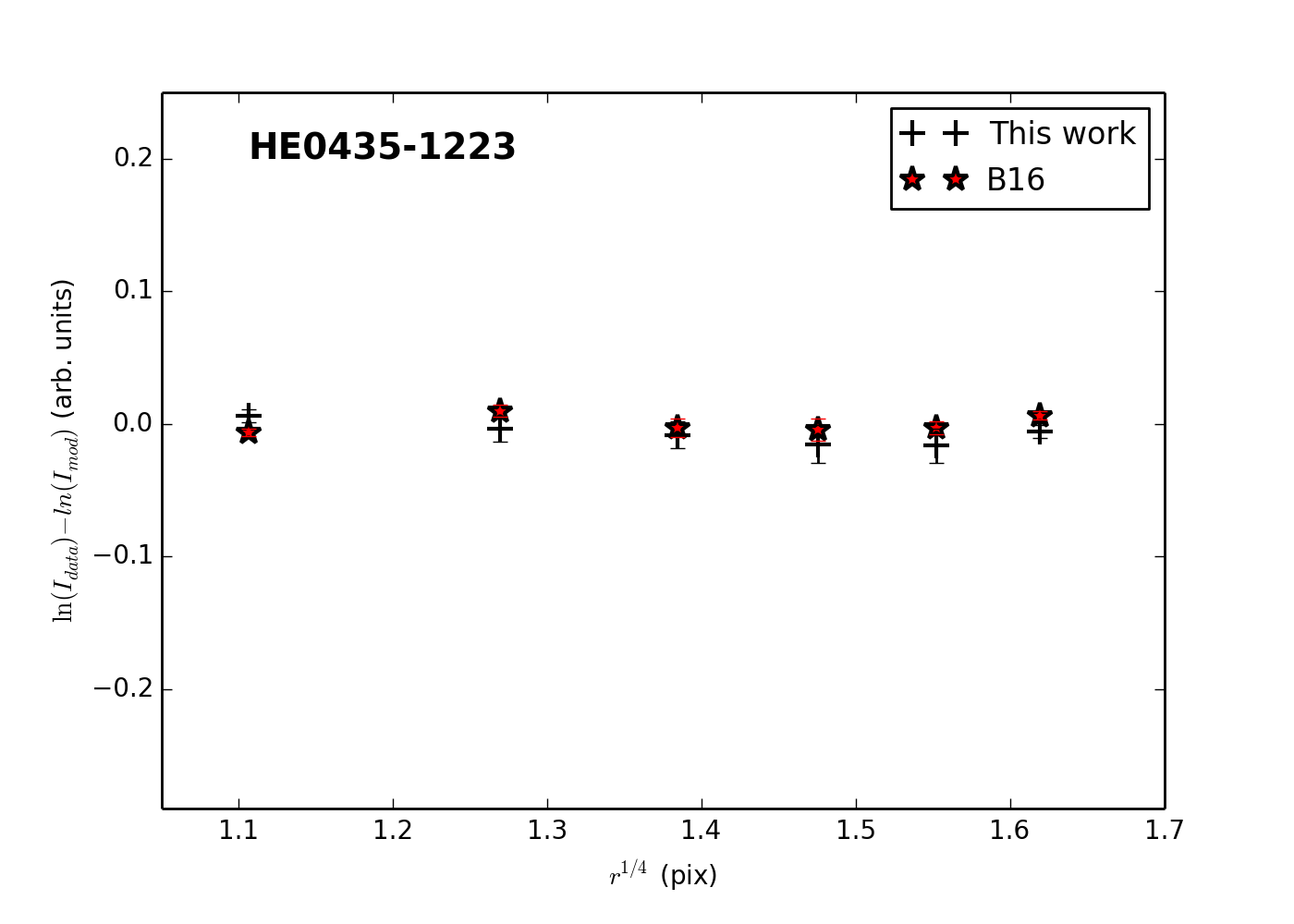} \\
\includegraphics[scale=0.45]{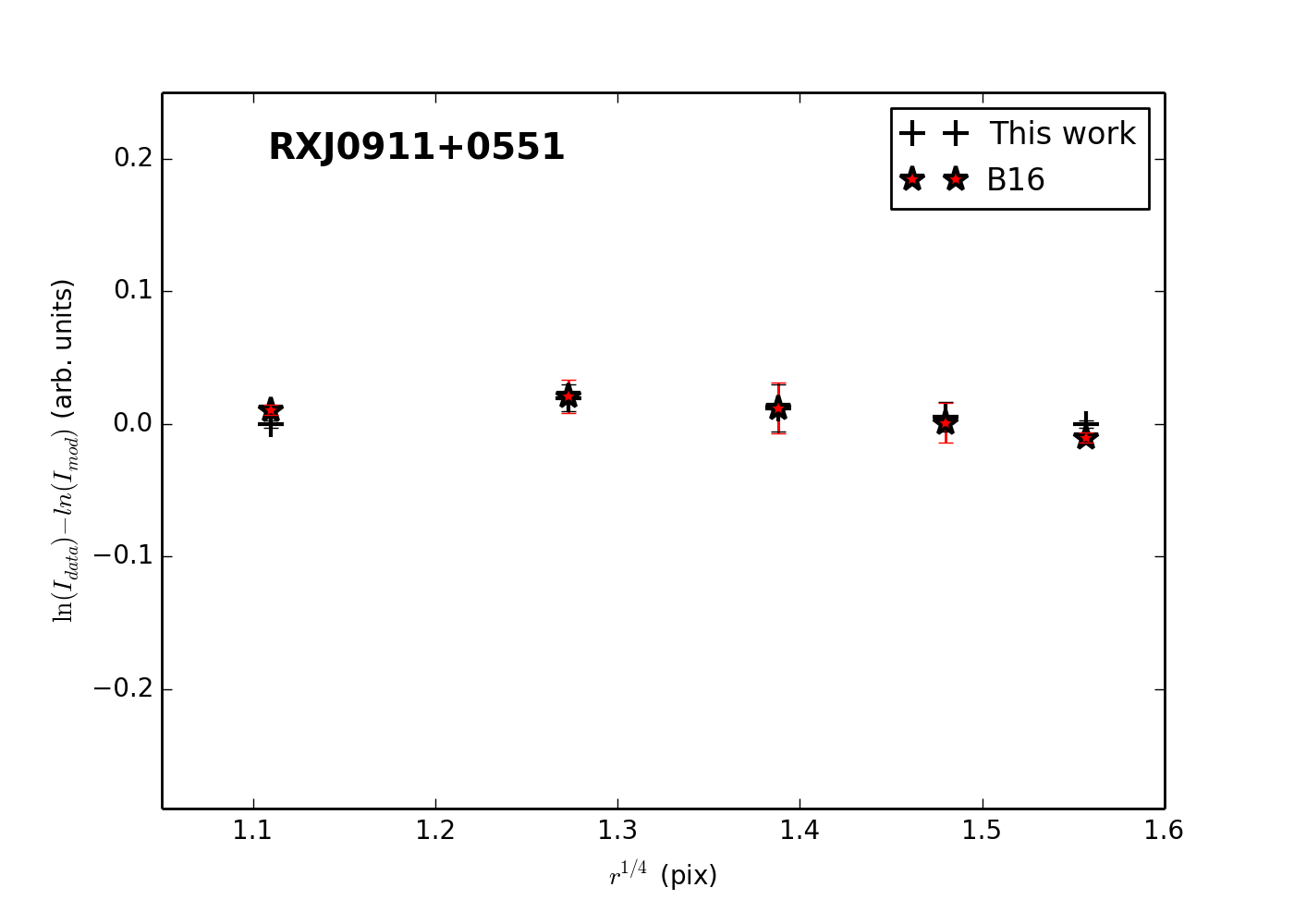} & \includegraphics[scale=0.45]{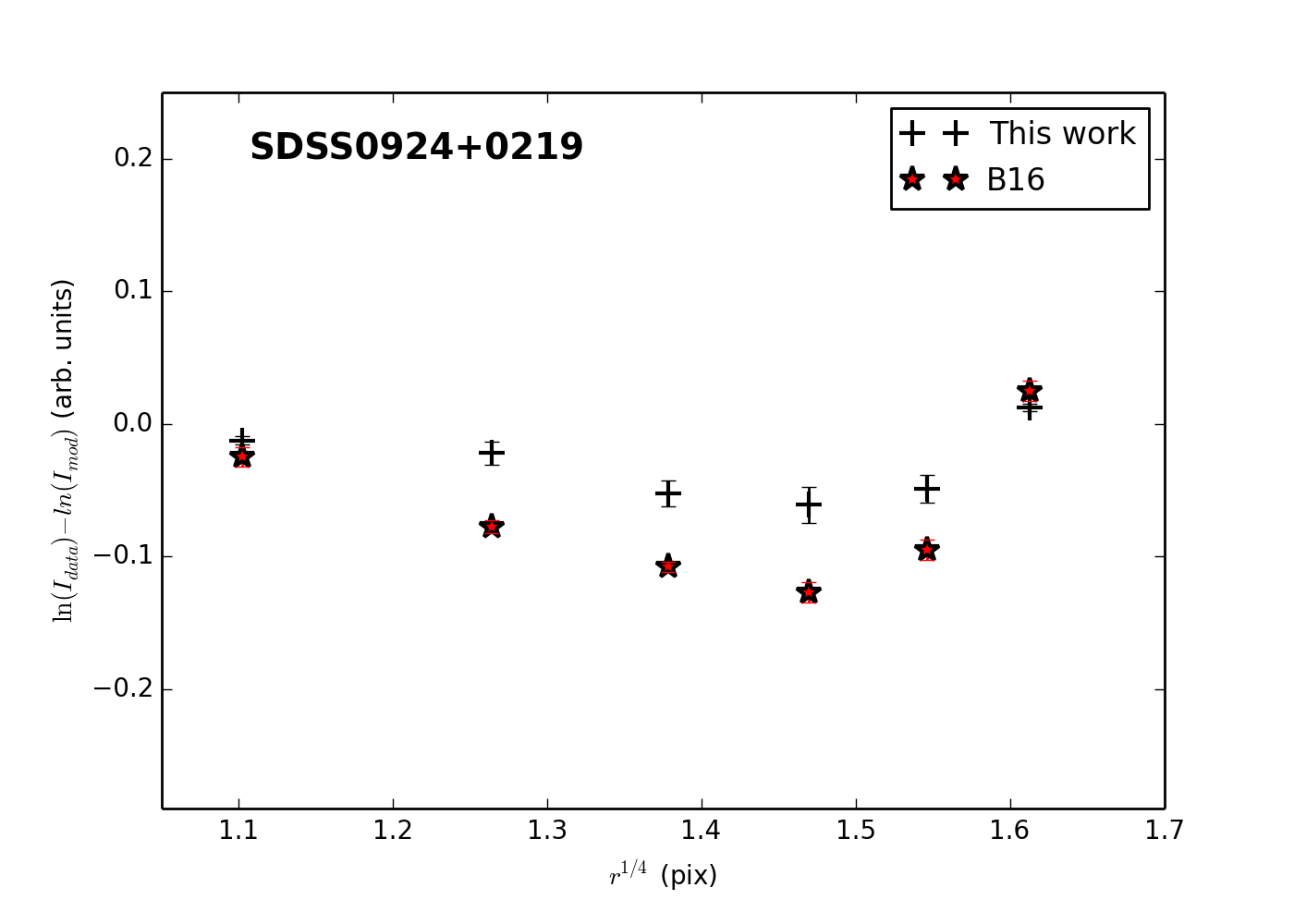} \\
\includegraphics[scale=0.45]{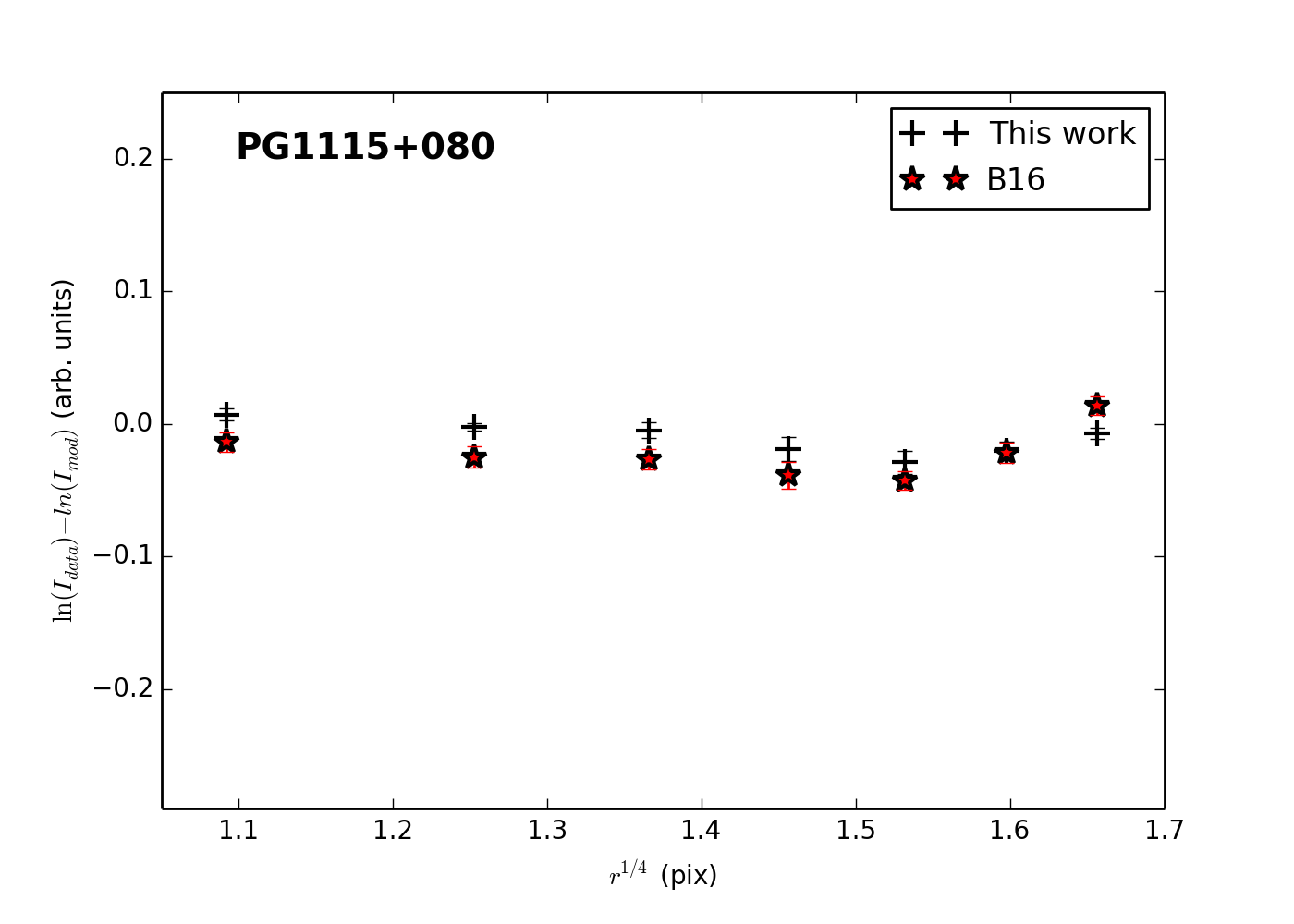} & \includegraphics[scale=0.45]{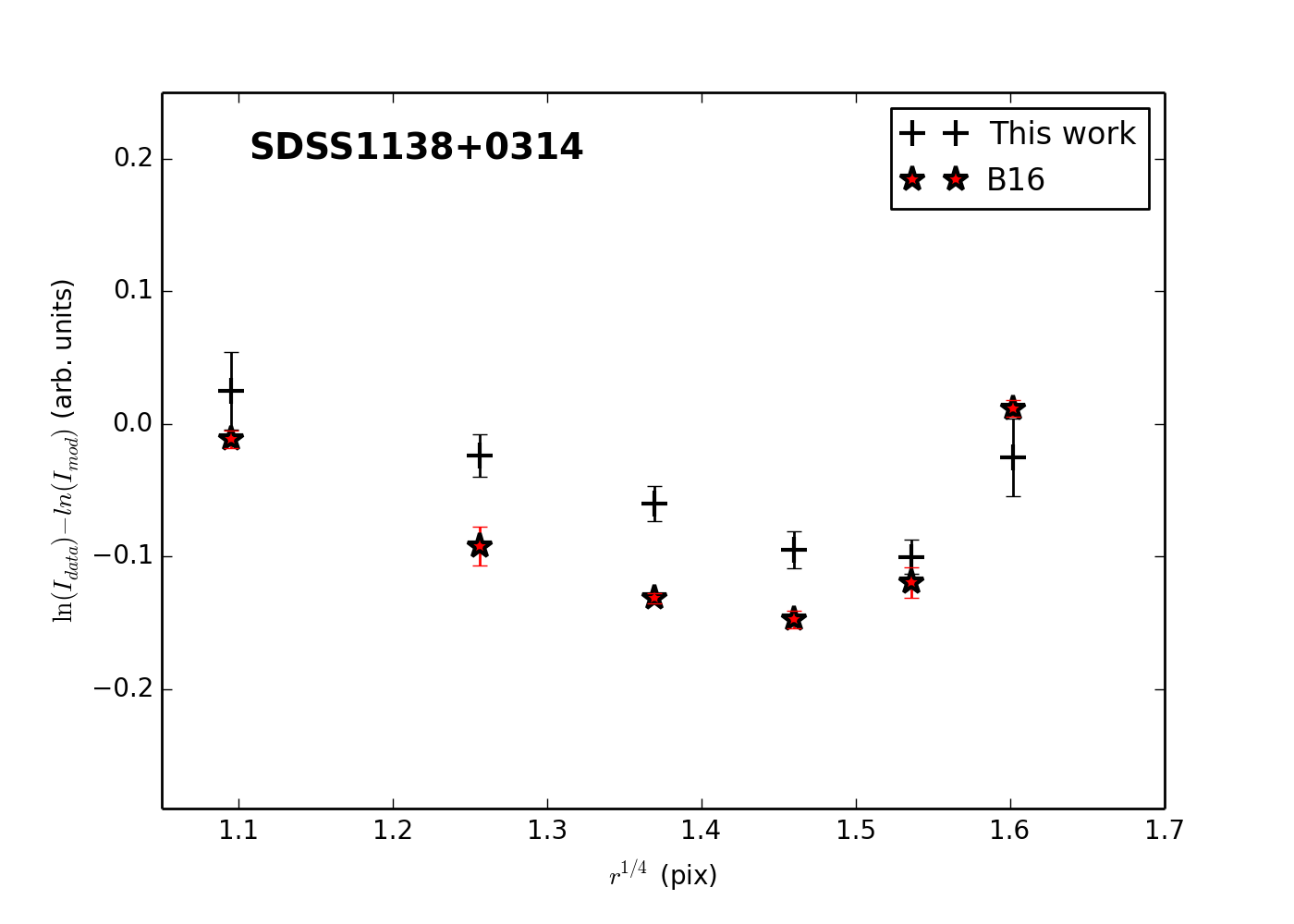} \\
\multicolumn{2}{c}{\includegraphics[scale=0.45]{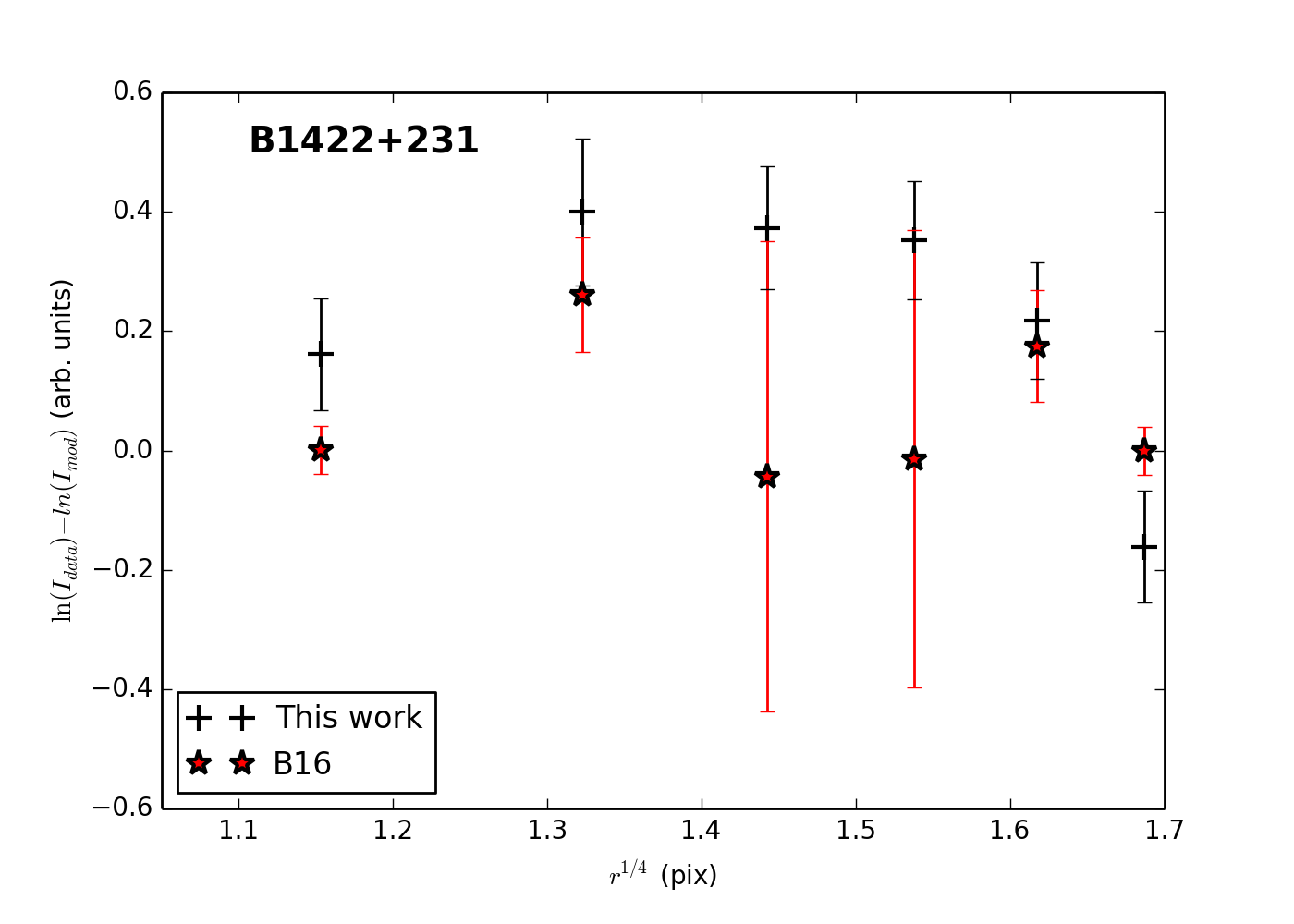}}
\end{tabular}
\label{fig_curv}
\end{figure*}

\indent In \citetalias{biernauxetal2016}, we performed a visual test to assess how well the de Vaucouleurs model represents the physical luminosity profiles of our lenses. The best model of each galaxy (the one appearing in Fig. \ref{fig_resi}) was convolved by the NIC2 PSF; its logarithmic radial profile $\ln I$ vs $r^{1/4}$ was computed and subtracted from that of the data. If any residual curvature remained, it would mean that $n \neq 4$, $n$ being the S\'ersic index. We arrived at the conclusion that in four out of seven cases, the systems displayed a significant upwards curvature, indicating that their S\'ersic index $n$ may be higher than four. After the arc subtraction and the change in the LRM, we re-conducted this test to see if it changed our previous conclusion. As can be seen in Fig \ref{fig_curv}, for cases where there was no curvature in \citetalias{biernauxetal2016} (HE0435-1223 and RXJ0911+0551), the results remains unchanged. On the other hand, for the four cases (MG0414+0534, SDSS0924+0219, SDSS1138+0314, and PG1115+080) where we saw a significant upwards curvature in \citetalias{biernauxetal2016}, the test now shows less or no curvature at all. The result is particularly visible for MG0414+0534, where no arc subtraction was performed. This not only provides validation to the de Vaucouleurs model hypothesis, but also to the new LRM validity criterion. B1422+231 goes from an inconclusive result to displaying a downwards curvature. However, on its data frames, one of the point sources appears close in projection to the galaxy. Therefore, separating signal from that point source and from the lens is highly uncertain. We address that problem by applying a weighting mask to B1422+231 with null weights for the ill pixels. However, in spite of that treatment, some artefacts from the sources subtraction may remain on the B1422+231 data frames. As can be seen in Tables \ref{tab_errreff} and \ref{tab_errflux}, the uncertainties coming from the positions and intensities of the sources are high for that system, and it has particularly large relative error bars. We therefore consider it less conclusive regarding the de Vaucouleurs profile test than the other systems.\\  

\indent By taking into account only one reasonable hypothesis (that the arc light distribution is radially symmetric), we can compute a more secure measurement of $r_{\rm{eff}}$. However, the arc subtraction comes with a few drawbacks: first, it may produce some residual noise or artefacts on individual pixels values. We take that into account by computing the error maps of the reconstructed arc image and by including it in the $r_{\rm{eff}}$ error bars. The main weakness of this arc subtraction process is that it requires a lot of human intervention at each step, mostly for verification purposes, and is therefore very time-consuming. Since each system presents a different arc, each one with its own specificities linked to its position, the point sources positions and the configuration of the system, building a quicker, more automated version of the arc subtraction may lead to the loss of some precision.\\

\indent The results of the $F_{\rm{H}}$ measurement are given in Table \ref{tab_k}, with their error bars and K-corrections. Since the $r_{\rm{E}}$ values depend very little on the lens modelling, and since our work includes careful error processing, $F_{\rm{H}}$ can be considered quite a robust estimate of the luminous flux in each lens. Together with the half-light radius measurement, these results make it possible to scale the distribution of ordinary matter in the lensing galaxies. Because lens modelling provides straightforward access to the total mass profile, this study could be the basis of a study of dark matter distribution in the lenses from our sample. We are currently addressing that question and the results are to be published in a future paper.\\

\section{Conclusions}\label{sec_ccl}

\indent We completed the work from \citetalias{biernauxetal2016} by adding the arc subtraction and improving the LRM. We obtained more secure measurements for $r_{\rm{eff}}$. Although the arc subtraction is time-consuming and requires human intervention at many steps, not taking the arc into account leads to an overestimate of $r_{\rm{eff}}$ of about 11\% in the case of our sample. We performed a highly detailed error calculation and obtained safe error bars for $r_{\rm{eff}}$ and $F_{\rm{H}}$. We have also verified that the de Vaucouleurs profile satisfactorily represents the physical light distribution of our sample elliptical galaxies. Eventually, our study yielded a high-quality characterisation of the quantity of luminous matter ($F_{\rm{H}}$) in the lenses from this sample. We intend to use these results to conduct a study of dark matter distribution in early-type galaxies, thus contributing to major debates in galaxy evolution and cosmology.  

\begin{acknowledgements} 
J. Biernaux acknowledges the support of the F.R.I.A. fund of the F.N.R.S. The authors are thankful to Dominique Sluse for his kind and valuable help.\\
\end{acknowledgements}

\bibliographystyle{aa}
\bibliography{sl_light}

\end{document}